\documentclass[twocolumn]{article}
\usepackage[utf8]{inputenc}
\usepackage{amsthm}
\usepackage{amssymb}
\usepackage{amsmath}
\usepackage{gensymb}
\usepackage{graphicx}
\usepackage{url}
\usepackage{framed}
\usepackage{float}
\usepackage{wrapfig}
\usepackage{multirow} 
\usepackage{tikz}
\usepackage{fullpage}
\usepackage[toc,page]{appendix}
\usepackage{physics}
\usepackage{listings}
\usepackage[font=small,labelfont=bf]{caption}
\usepackage{cite}
\usepackage{hyperref}
\hypersetup{
    colorlinks=true,
    linkcolor=blue,
    citecolor=blue,
    filecolor=magenta,      
    urlcolor=cyan,
}

\usetikzlibrary{arrows}
\usetikzlibrary{decorations.markings}
\newcommand{\angstrom}{\text{\normalfont\AA}}

\usepackage{csquotes}
\MakeOuterQuote{"}

\renewcommand{\vec}[1]{\boldsymbol{\mathbf{#1}}}

\usepackage{geometry}
\begin{document}

\title{\textbf{Upstream modes and antidots poison graphene quantum Hall effect}}

\author{N. Moreau$^1$, B. Brun$^{1}$, S. Somanchi$^2$, K. Watanabe$^3$, T. Taniguchi$^4$, C. Stampfer$^2$ \& B. Hackens$^1$*\\
\small{$^1$IMCN/NAPS, Universit\'e catholique de Louvain (UCLouvain), B-1348 Louvain-la-Neuve, Belgium} \\
\small{$^2$JARA-FIT and 2nd Institute of Physics - RWTH Aachen, Germany} \\
\small{$^3$ Research Center for Functional Materials,}\\
\small{National Institute for Materials Science, 1-1 Namiki, Japan} \\
\small{$^4$ International Center for Materials Nanoarchitectonics,}\\
\small{National Institute for Materials Science, 1-1 Namiki, Japan}\\
\small{*email: benoit.hackens@uclouvain.be}}

\date{}

\maketitle

\noindent\textbf{The quantum Hall effect is the seminal example of topological protection, as charge carriers are transmitted through one-dimensional edge channels where backscattering is prohibited. Graphene has made its marks as an exceptional platform to reveal new facets of this remarkable property. However, in conventional Hall bar geometries, topological protection of graphene edge channels is found regrettably less robust than in high mobility semi-conductors. Here, we explore graphene quantum Hall regime at the local scale, using a scanning gate microscope. We reveal the detrimental influence of antidots along the graphene edges, mediating backscattering towards upstream edge channels, hence triggering topological breakdown. Combined with simulations, our experimental results provide further insights into graphene quantum Hall channels vulnerability. In turn, this may ease future developments towards precise manipulation of topologically protected edge channels hosted in various types of two-dimensional crystals.}


\begin{figure*}[!ht]
\centering
\includegraphics[width=.7\linewidth]{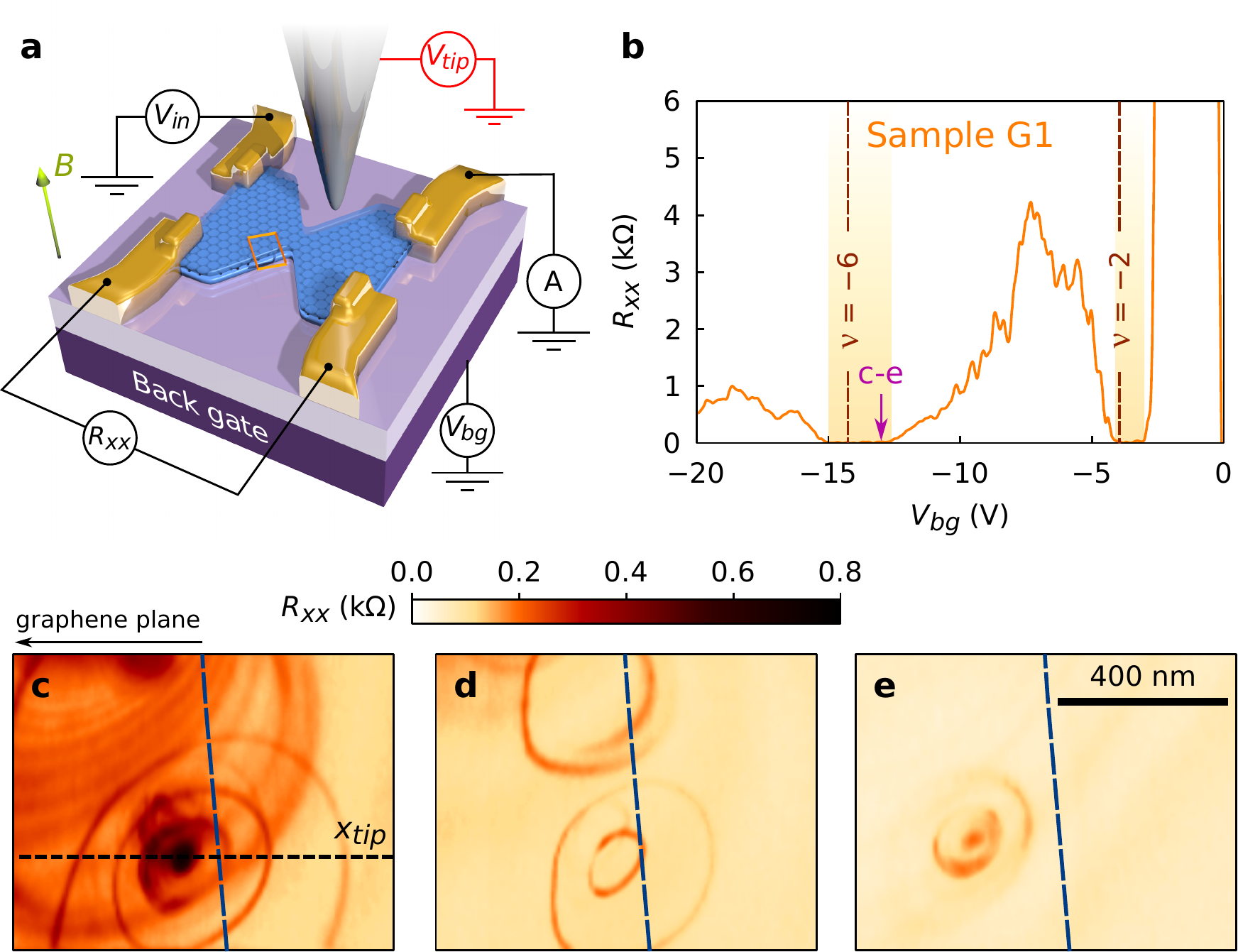}
\caption{\textbf{Imaging the topological protection break-down.} \textbf{a}, Schematic of the experimental setup. The biased tip can locally change the charge carriers density when applying the voltage $V_{tip}$ and is moved at a distance $d_{tip} \sim 70$ nm above the graphene plane. The global (bulk) charge carrier density in graphene is tuned by the back gate voltage $V_{bg}$. A magnetic field $B$ is applied perpendicularly to the graphene plane. \textbf{b}, Longitudinal resistance $R_{xx}$ as a function of $V_{bg}$, at $B = 10$ T, measured in sample G1. \textbf{c-e} SGM maps of $R_{xx}$ as a function of tip position. The scanning area is sketched by the orange rectangle in \textbf{a}, located $\sim$500 nm away from the constriction. The data are recorded with $V_{bg} = -13$V - arrow in \textbf{b} - and $V_{tip} = +3$V (\textbf{c}), $0$V (\textbf{d}) and $-6$V (\textbf{e}).}
\label{samples}
\end{figure*}

\begin{figure}[!ht]
\centering
\includegraphics[width=\linewidth]{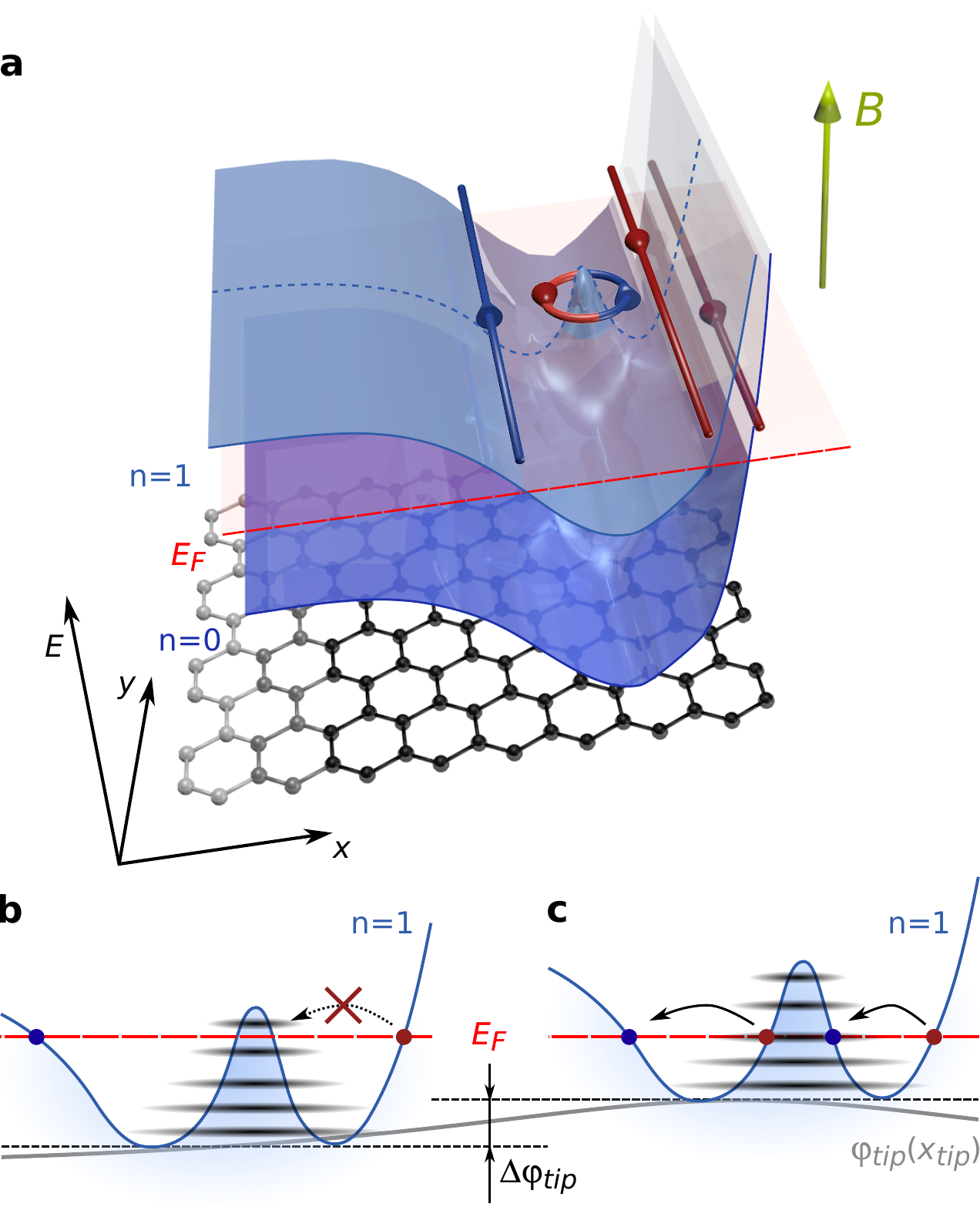}
\caption{\textbf{Artist' view of QHECs at graphene edge.} \textbf{a}, The two lowest LLs arising due to the perpendicular magnetic field (green arrow) are represented as blue semi-transparent surfaces. Because of electrostatics at the graphene edge (on the right side), they are bent and the $n = 1$ LL crosses twice the Fermi energy $E_F$ (red plane) yielding two downstream QHECs (in red) and to an upstream QHEC (in blue). An antidot is located between the counterpropagating QHECs and pins a QHEC island. \textbf{b},\textbf{c}, Line profile across the QHEC island (blue dotted line in A) for the $n = 1$ LL (electron charge carriers). Discrete energy levels are represented in black. The tip-induced potential $\varphi_{tip}(x_{tip})$ (gray line) tunes discrete energy levels positions with respect to $E_F$ when varying the tip position $x_{tip}$. When $E_F$ lies between two discrete energy levels, transport is not allowed via the QHEC island (\textbf{b}) whereas when a discrete energy level is aligned with $E_F$, charge carriers can tunnel between the counterpropagating QHECs (red and blue dots) through the QHEC island (\textbf{c}).}
\label{edge_structure}
\end{figure}


Quantum Hall edge channels (QHECs), formed as Landau levels (LLs) cross the Fermi energy near the borders of two-dimensional electronic systems (2DESs), are almost ideal one-dimensional systems, where quasiparticle scattering is topologically prohibited \cite{Halperin1982}. Substantial advances in the manipulation of QHECs in semiconductor-based 2DESs lead to envision new approaches in quantum computing \cite{Willett2013,Nakamura2019,Nakamura2020,Bartolomei2020,Baeuerle2018,Duprez2019} and open the way towards electron quantum optics \cite{Ji2003}. These breakthroughs require a robust topological protection of QHECs. 

Graphene, characterized by the massless nature of its charge carriers, offers even more promising perspectives in terms of QHECs manipulation, thanks to its rich spectrum of relativistic quantum Hall phenomena \cite{Goerbig2011}. In that framework, different strategies relying on QHEC propagation along p-n junctions have already been implemented in this material \cite{Rickhaus2015,Wei2017,Zimmermann2017,Bours2017,Jo2021}. However, the confinement of charge carriers at graphene borders appears much more difficult to control than in semiconductor-based 2DES, seriously impairing the topological protection of its QHECs. The explanation lies in different fundamental reasons, including the complex electrostatic screening of the back-gate potential related to the presence of fringing fields in most device layouts investigated up to now \cite{Silvestrov2008,Caridad2019}, and the difficulty to control defects at the borders of etched graphene \cite{Bischoff2015,Terres2016,Halbertal2017,Caridad2018}. The best proof of these detrimental influences is that some fractional quantum Hall signatures visible in extremely clean geometries were only observed in the case of edgeless device layouts such as the Corbino geometry \cite{Polshyn2018,Zeng2019}. 

Recently, local probe measurements \cite{Cui2016,Marguerite2019}, combined with theory \cite{Silvestrov2008}, led to a revision of the QHECs picture at graphene device edges. Instead of a single type of QHECs propagating along the border in clockwise or anticlockwise fashion as in semiconductor-based 2DES, the new proposed picture involves coexisting downstream and upstream QHECs separated by few-hundred-nm wide incompressible (\emph{i.e.} insulating) strips. Topologogical breakdown of graphene QHECs would therefore originate from the coupling between up- and downstream QHECs. This coupling has been revealed by Marguerite \textit{et al.} through scanning probe measurements \cite{Marguerite2019}: on one side, inelastic scattering were identified as a source of thermal dissipation along up- and downstream QHECs, with no incidence on transport, and on the other side, elastic tunneling was found to cause the coupling between these channels. However, the exact tunneling mechanism, and in particular a clear connection between scanning probe images and macroscopic transport properties, are still lacking.

\section*{Results}
\paragraph{Scanning gate microscopy in the quantum Hall regime.}
Here, we use scanning gate microscopy (SGM) to build a full microscopic picture of QHECs topological protection breakdown in graphene. For this purpose, we studied two devices (G1 and G2), consisting in 250 nm-wide encapsulated graphene constrictions as presented in Fig. \ref{samples}a. Figure 1b displays the longitudinal resistance $R_{xx}$ as a function of back gate voltage $V_{bg}$, showing fingerprints of the QH regime in graphene: $R_{xx}$ vanishes (orange-shaded boxes in Fig. \ref{samples}b) around the filling factors $\nu = \pm 4(n + 1/2)$, while it is maximal around $\nu = \pm 4n$ (the $n$-th LL is aligned with the Fermi energy - see supplementary section S1). 

In this work, we focus on the transition between the latter two regimes, where $R_{xx}$, while close to zero, exhibits fluctuations (see supplementary Fig. S6a), signatures of QH topological protection breakdown. Similar fluctuations have been evidenced in transport through constrictions defined in high mobility semiconductor-based 2DEG \cite{Ford1994,Goldman1995,Goldman2008,Hackens2010}. They have been ascribed to backscattering between QHECs propagating at opposite device edges, occurring through resonant tunneling via an antidot localized state. This mechanism is particularly effective when the antidot is located in the vicinity of the constriction where QHECs are brought in close proximity. 

The antidots locations in real space can be pinpointed thanks to SGM measurements. In SGM, local control over the potential landscape is achieved by electrically polarizing a sharp metallic tip moving in a plane parallel to the device surface. Recording simultaneously $R_{xx}$ as a function of tip position yields SGM maps. In the case of resonant tunneling between QHECs, a moving potential perturbation changes the resonance conditions, turning on and off QHECs backscattering. This yields circular features in SGM resistance maps, centered around the active antidot \cite{Hackens2010}.

In contrast with observations in semiconductor-based 2DEGs, centers of concentric SGM fringes are also located away from the constriction region of our graphene device. SGM images displayed in Figs. \ref{samples}c-e were obtained at a distance of ~500 nm from the constriction, at $V_{bg}=-13$ V, as indicated with an arrow in Fig. \ref{samples}b, \emph{i.e.} where the first deviations from $R_{xx}=0$ emerge, corresponding to the onset of the $\nu = -6$ QH state breakdown. SGM maps allow to pinpoint where the breakdown occurs: indeed, non-zero $R_{xx}$ regions draw sets of concentric rings centered close to the edges, whose number and position evolve with the tip polarization $V_{tip}$ (Figs. \ref{samples}c-e for sample G1 and supplementary Figs. S3b-f for sample G2). However, the observation of SGM contrast at large distance from the constriction (about 500 nm in Fig. \ref{samples}, and a few $\mu$m in suppl. Fig. S3) demonstrates that the constriction does not play a significant role here, which is counter-intuitive in the textbook framework of QH effect in conventional semiconductor-based 2DEGs. In this picture, counterpropagating QHECs run along opposite device edges, and are separated by an insulating bulk region much larger than the tip-induced perturbation. Away from the constriction, the edge states can only circumvent the perturbation and no tip-effect can be expected.

The key missing ingredient in the picture, allowing to solve the puzzling SGM signatures along the devices edges, is electrostatics. Indeed, as predicted by theory \cite{Silvestrov2008}, inhomogeneous screening of the back gate potential by graphene charge carriers leads to non-monotonic confining potential at the edges (see supplementary section S4 for further discussions about the effect that edge impurities could also have on this confining potential). Since LLs follow the same evolution as the potential, as schematically depicted in Fig. \ref{edge_structure}a, one then expects the presence of both up- and downstream QHECs along the same edge if the Fermi energy crosses twice the same LL. Tunneling between counterpropagating QHECs can be mediated by the presence of localized states associated with antidots, which pin circular QHECs "islands" in-between the QH channels (Fig. \ref{edge_structure}a). These antidots are at the origin of the characteristic concentric rings of non-zero $R_{xx}$ in Figs. \ref{samples}c-e. Note that these SGM signatures do not originate from a direct coupling of the counterpropagating QHECs induced by the tip potential alone: this would yield iso-resistance stripes following the edge topography \cite{Marguerite2019,Tomimatsu2020}. The absence of such stripes in SGM maps (Figs. \ref{samples}) testifies that the tip perturbation is small enough to avoid inducing direct backscattering.

\begin{figure*}[!ht]
\centering
\includegraphics[width=\linewidth]{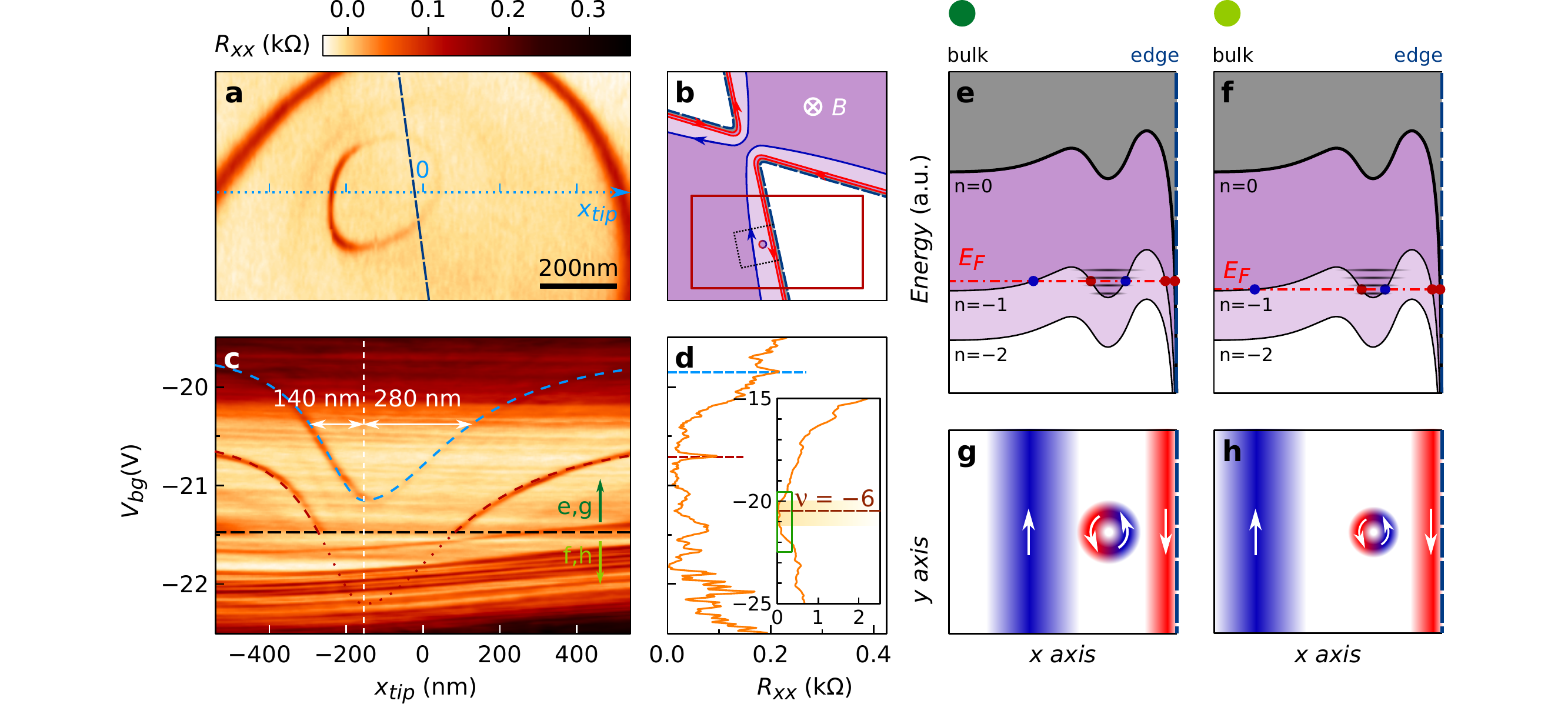}
\caption{\textbf{Coupling counterpropagating QHECs via an antidot.} The data are obtained in sample G1 for $V_{tip} = 0$ V and $B = 14$ T. \textbf{a}, SGM map obtained at $V_{bg} = -20.85$ V, by scanning the tip inside the red rectangle indicated in the schematic picture of the device sketched in \textbf{b}. \textbf{b}, The QHECs are represented by red (downstream) and blue (upstream) continuous lines, and dashed line delineate the constriction. \textbf{c}, $R_{xx}$ recorded as a function of $V_{bg}$ and the tip position $x_{tip}$, along the light blue dotted line in \textbf{a}. The resonances signatures (highlighted with the red and blue dashed lines) allow to measure the tip-induced potential variation at the QHEC island location as a function of $x_{tip}$. Blue and red dashed lines are fits obtained with two merged half-Lorentzian functions. Above graphene, the half width at half maximum is 140 nm whereas it is 280 nm when the tip is above the etched area. The black dashed line indicates the $V_{bg}$ limit beyond which one of the resonances disappears. \textbf{d}, Longitudinal resistance $R_{xx}$ as a function of $V_{bg}$ around $\nu = -6$ - zoom on the green rectangle of the inset. \textbf{e}, \textbf{f}, Schematics of the three lowest LLs, following the potential profile (thicker line) along $x_{tip}$-axis in map \textbf{c} for $V_{bg}>-21.5$ V (\textbf{e}) and $V_{bg}<-21.5$V (\textbf{f}). \textbf{g}, Schematic of the QHECs in real space, at the Fermi energy indicated by the red dash-dotted line in \textbf{e} (downstream in red and upstream in blue). The circular QHEC is pinned at the location of the antidot. \textbf{h}, Real space schematics of QHECs corresponding to Fermi energy indicated by the red dash-dotted line in \textbf{g}, where the upstream channel vanishes.}
\label{tip_shape}
\end{figure*}

\paragraph{Transport through antidots.}
Next we detail how the tip influences tunneling through such an antidot, whose electronic structure has been extensively studied in graphene with scanning tunneling microscopy \cite{Jung2011,Gutierrez2018,Walkup2020}. Antidots host discrete energy levels in the QH regime, whose positions are determined by size confinement in the resultant QHEC island on one hand (quantum contribution) and by Coulomb charging energy on the other hand (electrostatic contribution). A more in-depth discussion on the different contributions is given in supplementary section S2.2. Discrete energy levels are shifted under the tip-induced local modification of potential landscape, as sketched in Figs. \ref{edge_structure}b-c. The high $R_{xx}$ rings in Figs. \ref{samples}c-e are the loci of tip positions leading to an alignment between one of the antidot's discrete energy levels and the QHECs potential (Fig. \ref{edge_structure}c) whereas low $R_{xx}$ between the rings corresponds to Coulomb blockade \cite{Schnez2010,Garcia2013} (Fig. \ref{edge_structure}b). This picture is confirmed by the emergence of Coulomb diamonds in scanning gate spectroscopy \cite{Martins2013a}: applying a DC bias between source and drain allows to overcome Coulomb blockade as soon as the source-drain energy windows overlaps a localized state energy (see supplementary section S3). In this framework, the position of the antidot corresponds to the center of the Coulomb rings (at low $V_{tip}$, screening effects can however distort and shift Coulomb rings, as disccussed in supplementary section S2). Based on Figs. \ref{samples}c-e, we pinpoint antidots positions at a distance between 50 and 150 nm from sample G1 boundaries. This is in agreement with the estimated upstream QHEC position extracted from recent local probe results \cite{Marguerite2019,Cui2016}.

A fundamental question emerging at this point concerns the origin of the observed antidots. Atomic defects at the edges of graphene have often been invoked as source of perturbation for charge transport. However, if they were involved in the present case, it would remain to explain how they could yield potential landscapes similar to the one presented in Fig. \ref{edge_structure}a, with a potential extremum located 50 to 150 nm from the edge. More realistically, such potential landscape could originate from two known possible sources: (i) nanoscale random strain fluctuations, known to induce charge density inhomogeneities in graphene \cite{Couto2014} (ii) remote charged impurities in the dielectric hBN layer \cite{Onodera2019}. Both sources lead to local variations of Dirac point energies (typically $\sim 50-100$ meV at $B=0$ T, over typical distances $\sim 50-100$ nm \cite{Xue2011}), probably ubiquitous in all hBN/graphene/hBN heterostructures. While our experiment does not allow to discriminate between strain- or impurity-induced potential fluctuations, it  provides data on antidots distance from device borders, as well as on their spatial distribution along the borders of graphene devices : the typical distance between neighboring antidots is in the range 100-500 nm, from data in Figs. \ref{samples}c-e and S3, i.e. compatible with data from ref. \cite{Xue2011}]. Since the tip-induced potential perturbation extends beyond 500 nm, Coulomb rings originating from remote antidots can superimpose, as shown on Figs. \ref{samples}c-e and supplementary Figs. S3.

\begin{figure*}[!ht]
\centering
\includegraphics[width=\linewidth]{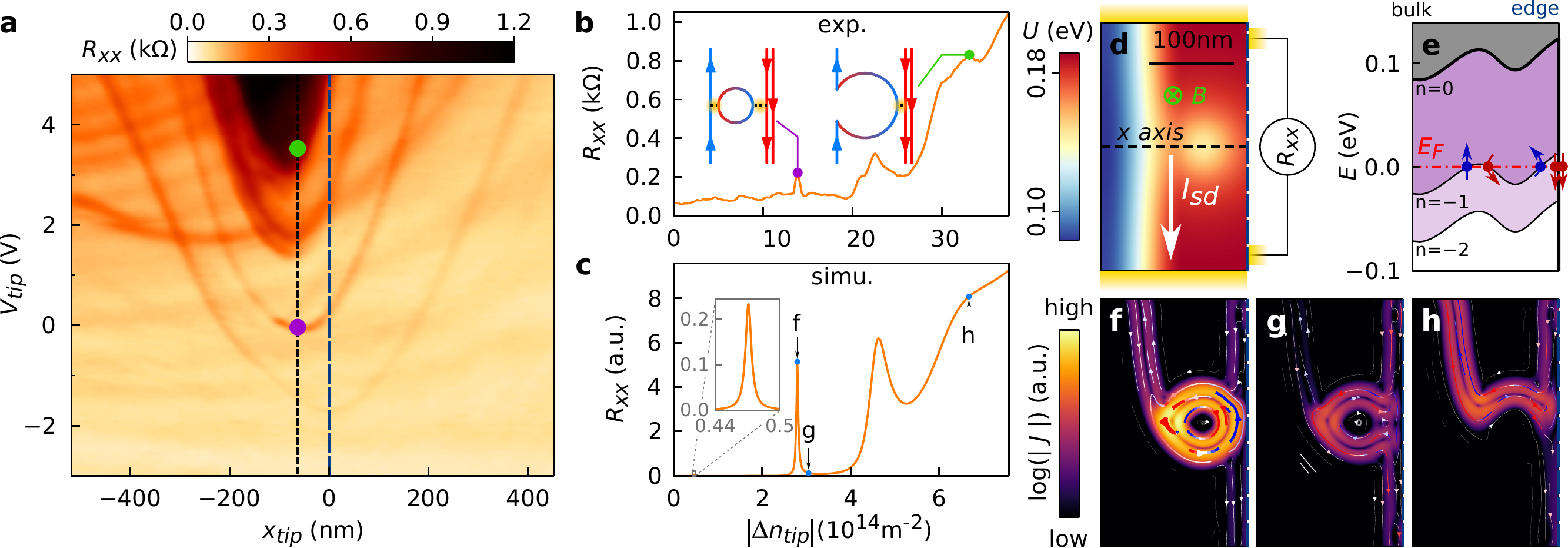}
\caption{\textbf{Tip-controlled tuning of transport through a QHEC island.} \textbf{a}, Evolution of $R_{xx}$ as a function of $V_{tip}$ (measured along the black dotted line in Fig. \ref{samples}c at $V_{bg} = -13$ V and $B = 12$ T). \textbf{b}, $R_{xx}$ profile for $x_{tip} \sim -60$ nm (tip on top of the antidot \emph{i.e.} along the black dotted line in \textbf{a}). $V_{tip}$ has been converted in the maximum tip-induced hole density decrease $|\Delta n_{tip}|$. \textbf{c}, Simulations of $R_{xx}$ as a function of $|\Delta n_{tip}|$ at the lower edge of $\nu=-6$ plateau, at $B = 12$ T. \textbf{d}, Scheme of the simulated system, with colors corresponding to the onsite potential landscape. The antidot corresponds to the circular region where the potential is lower, centered at $45$ nm from the graphene edge. The 4 leads required to compute $R_{xx}$ are represented in yellow. \textbf{e}, Profile of the three lowest LLs ($n = 0,-1,-2$) along the black dashed line in \textbf{d}. This graph is similar to Fig. \ref{tip_shape}f,g, except for the infinitely sharp confinement potential at the edge (right side of the figure) in the simulation, which yields two downstream QHECs (red straight arrows). \textbf{f-h}, Simulated maps of current density (JDOS) obtained for the three $\Delta n_{tip}$ values indicated with arrows in \textbf{c}. On top of the $R_{xx}$ peak (\textbf{f}), the JDOS around the antidot is maximal compared to the situation of zero $R_{xx}$ (\textbf{g}). The high JDOS in the antidot highlights that the resonance condition is reached. \textbf{h}, The region of finite $R_{xx}$ in \textbf{c} corresponds to direct backscattering of QHECs. The colored arrows indicate the direction of the local current density.}
\label{simus}
\end{figure*}

\paragraph{Back-gate and tip control of antidots}
The spatial locations of the antidots being unveiled, we now examine how their signatures emerge and evolve as a function of $V_{bg}$. For this purpose, we scan the tip across one of the antidots as indicated in Fig. \ref{tip_shape}a (the scan area in Fig. \ref{tip_shape}a corresponds to the red rectangle in Fig. \ref{tip_shape}b) and plot in Fig. \ref{tip_shape}c the SGM line profile as a function of $V_{bg}$ in the vicinity of $\nu = -6$ for a constant $V_{tip}$ (see supplementary section S2.3). It is well known from earlier SGM experiments on Coulomb blockaded islands that such a plot allows to infer the tip potential perturbation from the $V_{bg}$-shift of Coulomb blockade resonances \cite{Schnez2010,Pioda2004}. Coulomb resonances undergo a Lorentzian evolution, as shown by the fits in Fig. \ref{tip_shape}c, as expected for a tip-induced potential perturbation (see supplementary section S2.3). Examining Figs. \ref{tip_shape}a, c and d together, one can get the full picture of the fate of Coulomb resonances associated with antidots: Figs. \ref{tip_shape}c,d evidence that peaks identified by the red and blue dashed lines undergo a parallel evolution with the approaching tip perturbation, and are therefore associated with the same antidot, whose location is clearly identified in the SGM map in Fig. \ref{tip_shape}a. Importantly, the Coulomb resonances are also observed when the tip is far away from the device edges which means that the tunneling through the antidot is not necessarily triggered by the tip potential. Indeed, the Coulomb resonance signatures can be tuned by $V_{bg}$ as shown in Fig. \ref{tip_shape}d. 

A more intriguing behavior is also revealed for the resonance highlighted by the red dashed lines in Fig. \ref{tip_shape}c : below $V_{bg} < -21.5$ V (black dashed line), signatures of this resonance vanish. This $V_{bg}$ threshold is independent of $V_{tip}$ as demonstrated in supplementary Fig. S4. We propose the following picture to understand this phenomenon. Resonances are only visible provided that (i) a discrete state associated with an antidot is tunnel-coupled to up- and downstream QHECs as depicted in Fig. \ref{tip_shape}e,g and (ii) the upstream QHEC allows charge carriers to be sent back to the injection contact. Varying $V_{bg}$ has a strong influence on the position of the upstream QHEC (blue in Figs. \ref{tip_shape}g,h). As soon as the tunnel coupling becomes too small as illustrated in Figs. \ref{tip_shape}f,h or the upstream QHEC is no more connected to the injection contact, backscattering through the antidot is no longer effective and the resonance signature disappears. This data is crucial as it confirms the presence and the contribution of forward- and backward-propagating QH states at the device border. 

The coupling between the upstream QHEC and the injection contact is essential to understand the link between the QHECs structure and the filling factor deduced from transport measurements. Considering that this coupling is not perfect, the apparent filling factor is not defined by the bulk (dark purple in Fig. \ref{tip_shape}b) but rather by the incompressible region between the up- and downstream QHECs (light purple in Fig. \ref{tip_shape}b). In Figs. \ref{tip_shape}e,f, the filling factor therefore takes a value $\nu \sim -6$ even if the bulk filling factor is $-2$. We have further discussed the coupling between QHECs and the contacts in graphene samples in \cite{Moreau2021}.

Another way to tune the position and configuration of QHECs, but at the local scale, consists in varying both tip voltage and position. This is realized in Fig. \ref{simus}a showing the evolution of $R_{xx}$ when scanning the tip along the dashed line in Fig. \ref{samples}d and varying $V_{tip}$. The different visible resonances corresponding to the same antidot undergo parabolic evolution with $V_{tip}$ as expected for localized states \cite{Pioda2004}. At low $V_{tip}$, these resonances are separated by $R_{xx}\sim 0$ regions (corresponding to Coulomb blockade) while a finite $R_{xx}$ region (in dark in Fig. \ref{simus}a) is reached at larger positive $V_{tip}$. This evolution is also clearly visible in Fig. \ref{simus}b showing $R_{xx}$ versus the maximum tip-induced decrease in hole-density $|\Delta n_{tip}|$ deduced from $V_{tip}$ (see supplementary section S5), for a fixed $x_{tip}$ (with the tip on top of the antidot - black dotted line in Fig. \ref{simus}a). At lower tip perturbation, transport is determined by tunneling through the antidot as discussed above (left inset of Fig. \ref{simus}b). As the tip-induced perturbation increases, the antidot grows and merges with up- or downstream QHECs. The confinement of charge carriers in the antidot is then suppressed and the backscattering is only induced by the coupling between the counterpropagating QHECs, as depicted in the right inset of Fig. \ref{simus}b and further detailed in supplementary section S7.

\paragraph{Simulations}
Tight-binding simulations reproduce the observed phenomenology and provide further insights in the underlying physics through real space images of the local current density (JDOS) in the different backscattering regimes. Using the KWANT package \cite{Groth2014} (see supplementary section S8), we model one edge of the device as a 150 nm-wide graphene ribbon represented in Fig. \ref{simus}d where the colors correspond to the onsite potential landscape. In our simulations, we focus on a single side of the device, and neglect the bulk region contribution. The antidot potential is positioned close to the center of Fig. \ref{simus}d. In this geometrical configuration, counterpropagating QHECs (straight arrows in Fig. \ref{simus}e) encompass the QHEC island (curved arrows in Fig. \ref{simus}e) for the Fermi energy corresponding to the red dashed line of Fig. \ref{simus}e. The tip potential shifts the relative position of the LLs with respect to the Fermi energy, thereby tuning the distance and coupling between the QHECs and the antidot.

Noteworthy, we observe a striking qualitative correspondence between the measured (Fig. \ref{simus}b) and simulated (Fig. \ref{simus}c) longitudinal resistance as a function of $|\Delta n_{tip}|$ : at low $|\Delta n_{tip}|$, finite $R_{xx}$ peaks are separated by $R_{xx}\sim 0$ states and at larger $|\Delta n_{tip}|$, $R_{xx}$ remains finite. The $|\Delta n_{tip}|$ scale (distance between the peaks) depends mainly on the size of the considered antidot as well as on Coulomb interactions, not captured in our simulations. Since all the parameters vary among the antidots, the comparison between experimental and simulated typical $|\Delta n_{tip}|$ scales will remain qualitative. The sequence of JDOS maps shown in Figs. \ref{simus}f,g provides a real space illustration of the peaks' origin. Comparing Figs. \ref{simus}f and g, corresponding respectively to finite and zero $R_{xx}$ (see Fig. \ref{simus}c), we observe that, while in both cases the antidot is coupled to downstream QH channel (right of the figures), current through the antidot is significantly larger in the case of Fig. \ref{simus}f (as indicated by the brighter contrast in log scale at the antidot position). Coupling between up- and downstream QH channels is therefore much more efficient, yielding finite $R_{xx}$. At much higher $|\Delta n_{tip}|$ (Fig. \ref{simus}h), the JDOS map reveals that the raised antidot potential results in the merging of the antidot with the upstream QHEC, confirming the schematic picture sketched in the right inset of Fig. \ref{simus}b.

\section*{Discussion}
Put together, our data shed a new light on the combined role of electrostatics (fringing fields or charged impurities) and antidots at graphene edges in QH breakdown. Both ingredients are likely ubiquitous in most graphene-based heterostructures studied up to now, but with  variations in the importance of the different contributions. Indeed, fringing fields become much weaker when the gate is placed closer to graphene, for example when a graphite backgate is used below hBN. Furthermore, charged impurities at hBN etched edges depend on the etching recipe, and Dirac point inhomogeneites may be more or less pronounced depending on strain accumulated in the layers or on the quality of hBN. 

SGM data obtained at high magnetic field allow to get precise information on active antidot locations (distance from the border, and distribution along the border), putting constraints on their possible origin. The fine control over antidot size and coupling to QHECs provided by the tip and back-gate voltages was shown here to be the key to disentangle the complexity of the QH effect phenomenology in graphene. It allows to image and tune antidot-mediated QH effect breakdown, which constitutes a prerequisite towards advanced control and manipulation of QHECs in more complex devices such as QH interferometers. These findings are indeed relevant, for example, in the case of pn junction-based interferometers where semi-reflecting mirrors are defined at the edges \cite{Wei2017, Jo2021}. Noteworthy, the main outcome of this work, that full control over topological edge states in graphene will only be provided through meticulous  engineering of electrostatic landscape at device borders, can also be transposed to other types of 2D crystal-based devices hosting topologically-protected edge states. 

\section*{Materials and methods}

\begin{small}
\subsection*{Samples fabrication}
Sample G1, depicted in figure 1a, consists in a graphene flake encapsulated between two hBN layers (20 nm-thick for the top layer and 30 nm-thick for the bottom layer) using dry transfer techniques and deposited on a doped Si wafer covered by a 300 nm-thick SiO$_2$ layer. A 250 nm-wide constriction shape has been lithographically-defined, similarly to \cite{Terres2016}. The four contacts allow to measure the longitudinal resistance $R_{xx}$.

Sample G2, depicted in supplementary Fig. S2a, has been built with the same processes as sample G1 and with the same hBN layers thicknesses. The constriction has the same width. The major difference with sample G1 lies in the presence of six contacts, allowing to measure the Hall resistance $R_{xy}$ in addition to $R_{xx}$.

\subsection*{Measurements technique}
The sample has been anchored to the mixing chamber of a dilution refrigerator whose base temperature is 100 mK and a magnetic field $B$ up to 14 T has been applied perpendicularly to the graphene plane. Electrical signals have been recorded using a classical lock-in technique at a frequency of 77 Hz. The longitudinal resistance is obtained from a four probe measurement to avoid the contribution from contacts resistance. Charge carriers type and density can be tuned by changing the back-gate voltage $V_{bg}$. 

The local gate used for SGM characterization consists in a commercial metal-coated AFM tip glued on a tuning fork whose resonance frequency is $f\sim 32$ kHz. The tip is electrically contacted so that a voltage $V_{tip}$ can be applied on it. The tip can be moved in $x, y, z$ directions thanks to piezo scanners. After scanning the surface in topography mode, the distance $d_{tip}$ between the tip and the graphene plane can be fixed. Applying the bias $V_{tip}$ introduces an electrostatic perturbation for conduction electrons. The conductance can then be recorded for each tip position, yielding a SGM map.

\subsection*{Simulations}
Tight-binding simulations have been performed using the KWANT package \cite{Groth2014}. We modeled one edge of the device, neglecting the bulk region contribution (see supplementary Fig. S8), as a 150 nm-wide graphene ribbon represented in Fig. \ref{simus}d where the colors correspond to the onsite potential landscape. This potential is asymmetric along the $x$-axis, resulting in the spatial profile for the LLs shown as black lines in Fig. \ref{simus}e. Their shape matches the qualitative picture given in Figs. \ref{tip_shape}e,g for the energy levels' evolution close to the edge of the graphene device. Note that the confinement is infinitely sharp in the simulation at the device border (right side of Figs. \ref{simus}d-e), contrary to the smoother evolution schematically depicted in Fig. \ref{tip_shape}e and g, without consequence on the qualitative correspondence between simulation and experimental results. Finally, the antidot potential has been modeled by a Gaussian function and is positioned at 45 nm from the edge. 

To decrease computation time, a scaling factor of four, without incidence on the output results, was applied to the real lattice parameter of graphene (the interatomic distance is $a = 4\times 1.42 ~ \angstrom$ and the hopping parameter is $t = 2.7/4$ eV). More details on simulations are available in supplementary section S8.

\subsection*{Data availability}
The tight-binding code used to produce the simulations presented in this article are available at \url{https://github.com/moreaunic/Simulations_graphene_QHE}

The experimental data that support the findings of this study are available from the corresponding authors on request.

\bibliographystyle{naturemag}
\bibliography{biblio.bib}{}

\section*{Acknowledgments}
The authors thank Vincent Bayot and Markus Morgenstern for their valuable comments and corrections. N.M. thanks Sébastien Toussaint for enriching discussions. B.B. thanks Julien Renard for sharing unpublished data. This work was partly funded by the Federation Wallonie-Bruxelles through the ARC Grant No. 16/21-077, by the F.R.S-FNRS through the Grant No. J008019F, and from the European Union’s Horizon 2020 Research and Innovation program (Core 1 No. 696656 and Core 2 No. 785219). This work was also partly supported by the FLAG-ERA grant TATTOOS, through F.R.S.-FNRS PINT-MULTI grant No. R 8010.19. Computational resources have been provided by the Consortium des Equipements de Calcul Intensif (CECI), funded by the Fonds de la Recherche Scientifique de Belgique (F.R.S.-FNRS) under Grant No. 2.5020.11 and by the Walloon Region. B.B. (research assistant), B.H. (research associate) and N.M. (FRIA fellowship) acknowledge financial support from the F.R.S.-FNRS of Belgium. Support by the Helmholtz Nanoelectronic Facility (HNF), the EU ITN SPINOGRAPH and the DFG (SPP-1459) is gratefully acknowledged. K.W. and T.T. acknowledge support from the Elemental Strategy Initiative conducted by the MEXT,
Japan ,Grant Number JPMXP0112101001, JSPS KAKENHI Grant Numbers JP20H00354 and the CREST(JPMJCR15F3), JST.

\section*{Author contributions}
N.M performed the experimental measurements, with the assistance of B.B. N.M. analyzed the data. N.M. performed the tight-binding simulations. S.S. fabricated the samples. K.W. and T.T. synthesized the h-BN crystals. N.M. and B.H. wrote the paper. B.H. and C.S. supervised the collaboration. All authors discussed the results and commented on the manuscript.

\end{small}

\onecolumn

\newgeometry{left=30mm, right=30mm, top=25mm, bottom = 35mm, }
\renewcommand{\thefigure}{S\arabic{figure}}
\renewcommand{\thesection}{S\arabic{section}}
\setcounter{figure}{0}
\onecolumn
\newpage

\vspace*{20mm}

\begin{center}
\textbf{\LARGE{Upstream modes and antidots poison graphene quantum Hall effect}}\\ 
\vspace*{5mm}
\textbf{\LARGE{- Supplementary materials -}}
\end{center}

\vspace*{10mm}
\section{Fan diagrams and determination of the lever-arm}\label{sec:lever-arm}

\begin{figure*}[!ht]
\centering
\includegraphics[width=.8\linewidth]{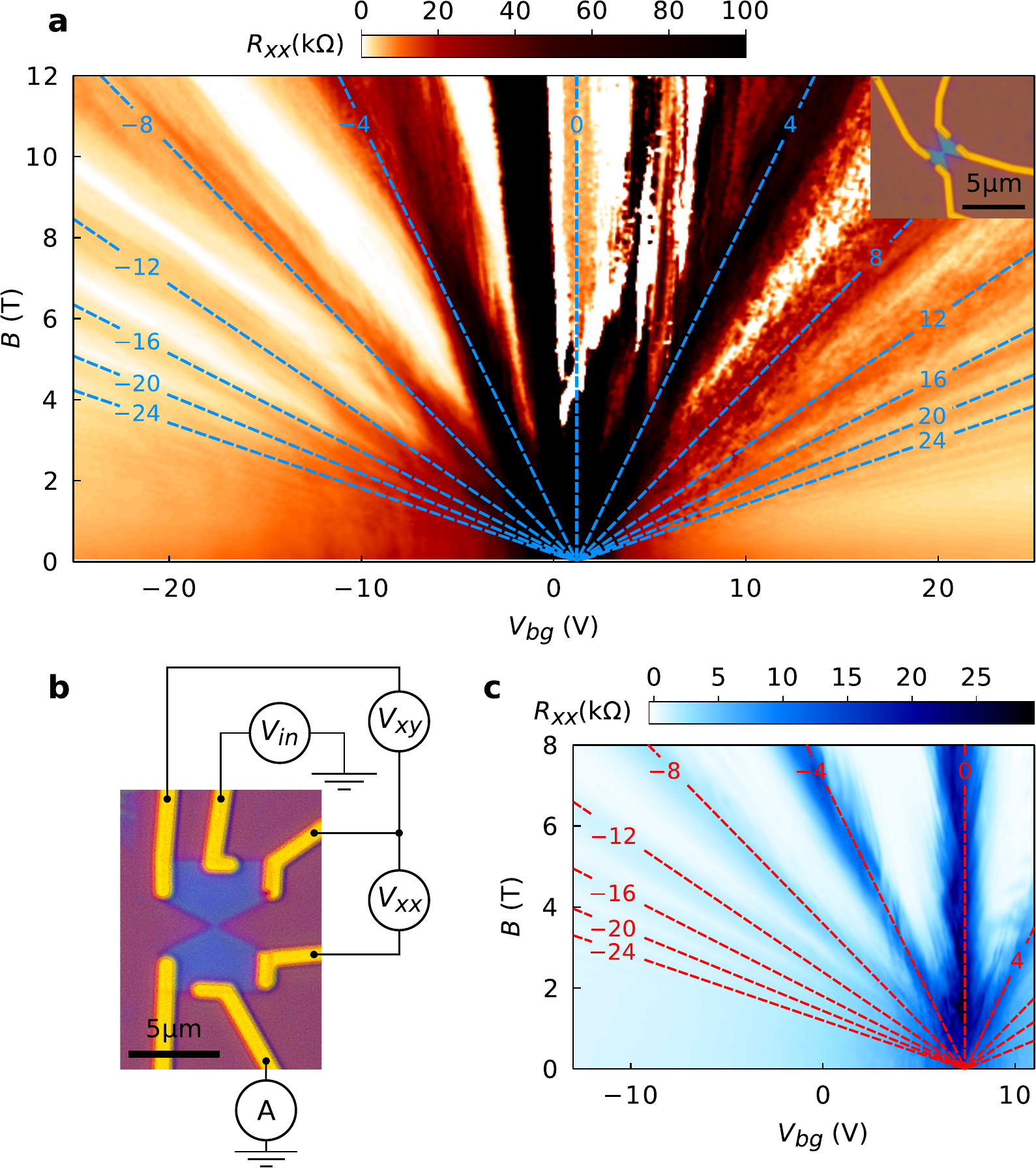}
\caption{\textbf{Fan diagrams for samples G1 and G2} \textbf{a}, $R_{xx}$ map as a function of $V_{bg}$ and $B$ for sample G1. The dashed lines follow the couples $(V_{bg}, B)$ for which, in a textbook representation of quantum Hall effect, a bulk Landau level is aligned with the Fermi energy ($R_{xx}$ exhibits local maxima along these lines). These maxima coincide with the filling factors $\nu$ that label each line. The inset shows an optical picture of sample G1. The BN-encapsulated graphene device has a blue color and the 4 gold contacts have a yellow color. \textbf{b}, Optical picture of sample G2. The 6 contacts of this sample allow to measure both longitudinal resistance $R_{xx}$ and Hall conductance $G_{xy}$. \textbf{c}, $R_{xx}$ map as a function of $V_{bg}$ and $B$ for sample G2. As in \textbf{a}, the dashed lines are labeled by the filling factor.}
\label{Fan_diagrams}
\end{figure*}

In this work, we studied two samples (G1 and G2). Both consist in hBN-encapsulated monolayer graphene etched in a constriction-shaped geometry as shown with optical microscopy photographies in the inset of Fig. \ref{Fan_diagrams}a for sample G1 and in Fig. \ref{Fan_diagrams}b for sample G2. The charge carriers density $n$ in graphene is tuned via the back gate voltage $V_{bg}$ applied on the degenerately doped silicon substrate separated from graphene by a 300 nm-thick layer of silicon oxide and the 20 nm-thick bottom hBN. The lever arm $\alpha = n/V_{bg}$ can be extracted for both samples by recording longitudinal resistance $R_{xx}$ while varying $V_{bg}$ (the Fermi energy $E_F$ is shifted) and the magnetic field $B$ (the energy gap between the Landau levels (LLs) depends on $B$). In a textbook representation of quantum Hall effect, backscattering is maximized when a $E_F$ is aligned with one of the LL. It results in strips of local $R_{xx}$ maxima in the $V_{bg}-B$ map located along the straight lines
\begin{equation}\label{lever_arm}
B_{n} = \dfrac{\pi \hbar}{2en}\alpha(V_{bg}-V_{bg,0})
\end{equation}
where $n$ are the LLs labels (negative for holes and positive for electrons) and $V_{bg,0}$ is the back gate voltage corresponding to charge neutrality ($+1.2$ V for sample G1 and $+7.4$ V for sample G2). Fits to the $R_{xx}$ maxima strips were performed for both samples (dotted lines in Fig. \ref{Fan_diagrams}a for sample G1 and Fig. \ref{Fan_diagrams}c for sample G2) and allowed to determine the values of $\alpha$ about $9.4\times 10^{14}$ m$^{-2}$V$^{-1}$.

\section{Effect of the tip potential screening in the SGM signatures}\label{screening}

In this section, we first present the SGM measurements obtained with sample G2. Peculiar SGM signatures appear in this device at low tip voltage, that we attribute to screening effects of the tip potential by the back gate. Then, we detail the resonance condition enabling the coupling between counterpropagating quantum Hall edge channels (QHECs) through an antidot. Finally, we further discuss the interplay between the tip, the back gate and the graphene plane and their mutual screening. It has important implications in the characterization of the tip-induced potential.

\subsection{SGM measurement on sample G2}\label{SGM_2_sec}
\begin{figure*}[!ht]
\centering
\includegraphics[width=\linewidth]{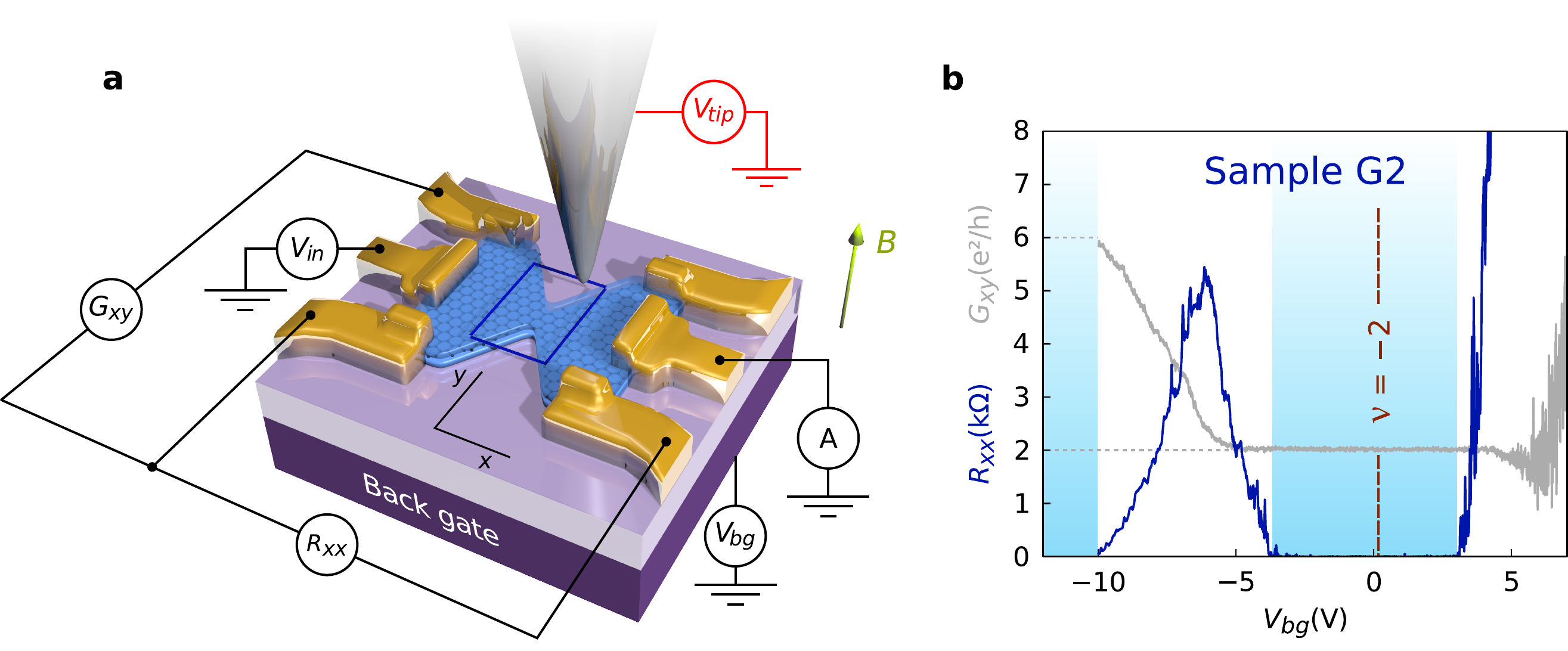}
\caption{\textbf{Experimental setup for sample G2} \textbf{a}, Schematic of sample G2. The biased tip can locally change the electron density when applying the voltage $V_{tip}$ and is moved at a distance $d_{tip} \sim 50$ nm above the encapsulated graphene constriction. The global electron density is tuned by the back gate voltage $V_{bg}$. A perpendicular magnetic field $B$ is applied perpendicularly to the graphene plane. \textbf{b}, Transport measurements (the longitudinal resistance $R_{xx}$ in blue and the Hall conductance $G_{xy}$ in gray) obtained for $B = 14$ T. The light blue stripes indicate the region of zero $R_{xx}$, associated with conductance plateau at $2e^2/h$ around $\nu = -2$ and $6e^2/h$ around $\nu = -6$.}
\label{Device_2}
\end{figure*}

All experimental data presented in the manuscript were obtained on sample G1. In this section, we present the SGM measurements obtained on our second device (sample G2). The data exhibit the same features as presented in the manuscript. A schematic of the experimental setup used for sample G2 is shown in Fig. \ref{Device_2}a. The longitudinal resistance $R_{xx}$ as well as the Hall conductance $G_{xy}$ recorded as a function of back gate voltage $V_{bg}$ under a magnetic field of 14 T are presented in the graphics of Fig. \ref{Device_2}b. 

\begin{figure*}[!ht]
\centering
\includegraphics[width=.8\linewidth]{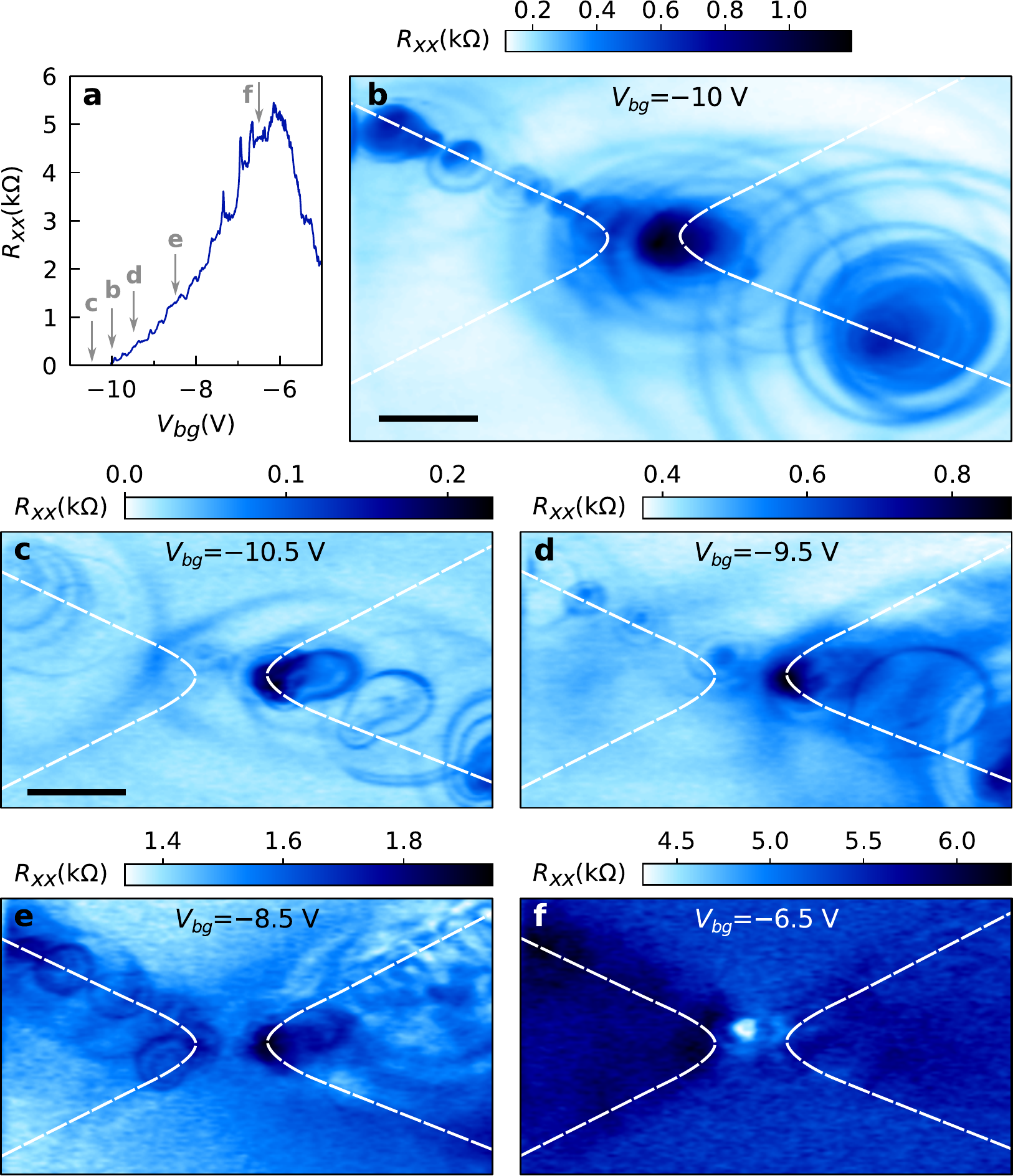}
\caption{\textbf{Imaging the break-down of topological protection in sample G2.} \textbf{a}, Zoom on Fig. \ref{Device_2} between $\nu = -6$ and $\nu = -2$ corresponding to the loss of topological protection (non-zero $R_{xx}$). \textbf{b-f}, SGM maps obtained at 14 T and $V_{tip} = +3$ V (\textbf{b}) and $V_{tip} = 0$ V (\textbf{c}-\textbf{f}). The $V_{bg}$ values used for each map are indicated with the arrows in \textbf{a}. The rings of higher $R_{xx}$ are the loci of tip positions leading to a resonance between an antidot localized state and the counterpropagating QHECs and hence the beakdown of topological protection of these channels.}
\label{SGM_2}
\end{figure*}

Figure \ref{SGM_2} presents several SGM maps obtained at 14 T by scanning the tip at $\sim 55$ nm above the graphene plane within the scan area sketched in Fig. \ref{Device_2}a. As for the manuscript's data, we explored the transition between $\nu = -6$ and $\nu = -2$ (Fig. \ref{SGM_2}a) where breakdown of QHECs topological protection occurs. In these conditions, the same fingerprints of high resistance rings along the device edges emerge in sample G2, highlighting the presence of antidots as the cause of topological breakdown. In the manuscript (sample G1), we explain that the rings are centered above the antidots, at a distance between 50 and 150 nm from device borders, in agreement with the QHECs region width reported in literature \cite{Cui2016}. In Fig. \ref{SGM_2}b, however, the rings are centered on the device's borders. In figures \ref{SGM_2}c-f, SGM signatures are even not circular and exhibit peer shapes.

We explain these surprising observations by two mechanisms: 
\begin{itemize}
\item[•] Metallic gates and graphene are known to screen the tip potential, leading to shifts and deformations of the SGM signatures. In particular, Schnez et al. studied a graphene quantum dot with SGM and observed that the circular signatures in SGM maps, caused by Coulomb blockade, were shifted because of the side gates used to control the charging state of the dot \cite{Schnez2011}. We interpret our observed shifts and deformation of high $R_{xx}$ rings as screening of the tip-induced potential by the back gate. We are comforted in our interpretation by the fact that deformations of the high $R_{xx}$ rings are stronger at low $V_{tip}$ (Figs. \ref{SGM_2}c-f with $V_{tip} = 0$ V) than at high $V_{tip}$ (Fig. \ref{SGM_2}b with $V_{tip} = +3$ V) where rings are circular. Screening effects will be discussed in more details in section \ref{tip_screening}.

\item[•] There is an uncertainty on the position of sample G2 edges coming from the fact that they were determined by electrostatic force microscopy. This technique consists in mapping the force applied on the tip, related to the electrostatic interactions with the substrate. When the tip lies above graphene or above the remote polarized back gate, both grounded metallic plane but located at different distances from the tip, the resulting electrostatic interaction in different. The transition of electrostatic environment when crossing the edge result in a smooth signal when the tip is 55 nm above the graphene plane and hence an uncertainty of about 100 nm in our case. In sample G1, edges positions were determined more conventionally by atomic force microscopy (AFM) and the sharp transition of the measured height at the edges decreases the uncertainty on their position.
\end{itemize} 

\subsection{Resonance condition}
Here, we detail the resonance condition of an antidot, whose electronic structure has been extensively studied in two recent papers \cite{Gutierrez2018, Walkup2020} (note that in these cases antidots were created by introducing charges in the hBN substrate whereas they preexist in our samples). When the antidot charging energy is equal for $N$ and $N+1$ charge carriers, charge-discharge events can occur. In the QH regime, the antidot charging energy is given by \cite{Hanson2007, Rosenow2007, Ihnatsenka2009, Ofek2010}
\begin{equation}\label{charging}
U(N) = \dfrac{1}{2C_{\Sigma}}\left( qN - qf_c\dfrac{B\delta A}{\phi_0} + C_{tip}V_{tip} + C_{bg}V_{bg}\right)^2 + \sum_{n=1}^{N}E_n
\end{equation} 
where $C_{\Sigma}$ is the antidot capacitance, $q$ the electron ($-e$) or hole ($+e$) charge, $f_c$ the number of QHECs around the antidot, $\delta A$ the area variation of the antidot, $C_{tip}V_{tip}$ and $C_{bg}V_{bg}$ the charge imbalance induced by the tip and the back-gate voltage respectively and $E_n$ the quantum resonant energy levels. By moving the biased tip near the antidot, the capacitance term $C_{tip}$, the area $\delta A$ as well as the quantum terms $E_{n}$ in Eq. \ref{charging} vary according to the tip position $\vec{r}_{tip}$. The Coulomb rings in Figs. 1c-d and Figs. \textbf{S3}b-f are the loci of $\vec{r}_{tip}$ for which the resonance condition $U(N) = U(N+1)$ is reached, when the system is at equilibrium. As will be discussed in section \textbf{S7}, the quantum terms are small compared to the Coulomb contribution. 

\subsection{Characterization of the tip-induced perturbation}\label{tip_screening}
In Fig. 3b of the manuscript, we showed that mapping $R_{xx}$ vs $V_{bg}$ and $y_{tip}$ (along a line passing above an antidot) yields a direct image of the tip-induced potential felt by the antidot for a given $V_{tip}$\footnote{The workfunction difference between the tip and graphene plane should be added to the tip voltage so that around $+3$ V should be added to the effective $V_{tip}$ used in Eq. \ref{charging} \cite{Yu2009}}. Indeed, the induced potential was given by the shift in the Coulomb resonance signatures ($R_{xx}$ peaks) along $V_{bg}$ corresponding to the alignment of one of the antidot resonance levels with the Fermi energy. We then used Lorentzian functions to fit the induced potential and the function turns out to decay faster when the tip is above the graphene plane, which is expected from the screening of the graphene charge carriers. Here, we explore in more details the influence of screening on the tip-induced potential.

\begin{figure*}[!ht]
\centering
\includegraphics[width=.85\linewidth]{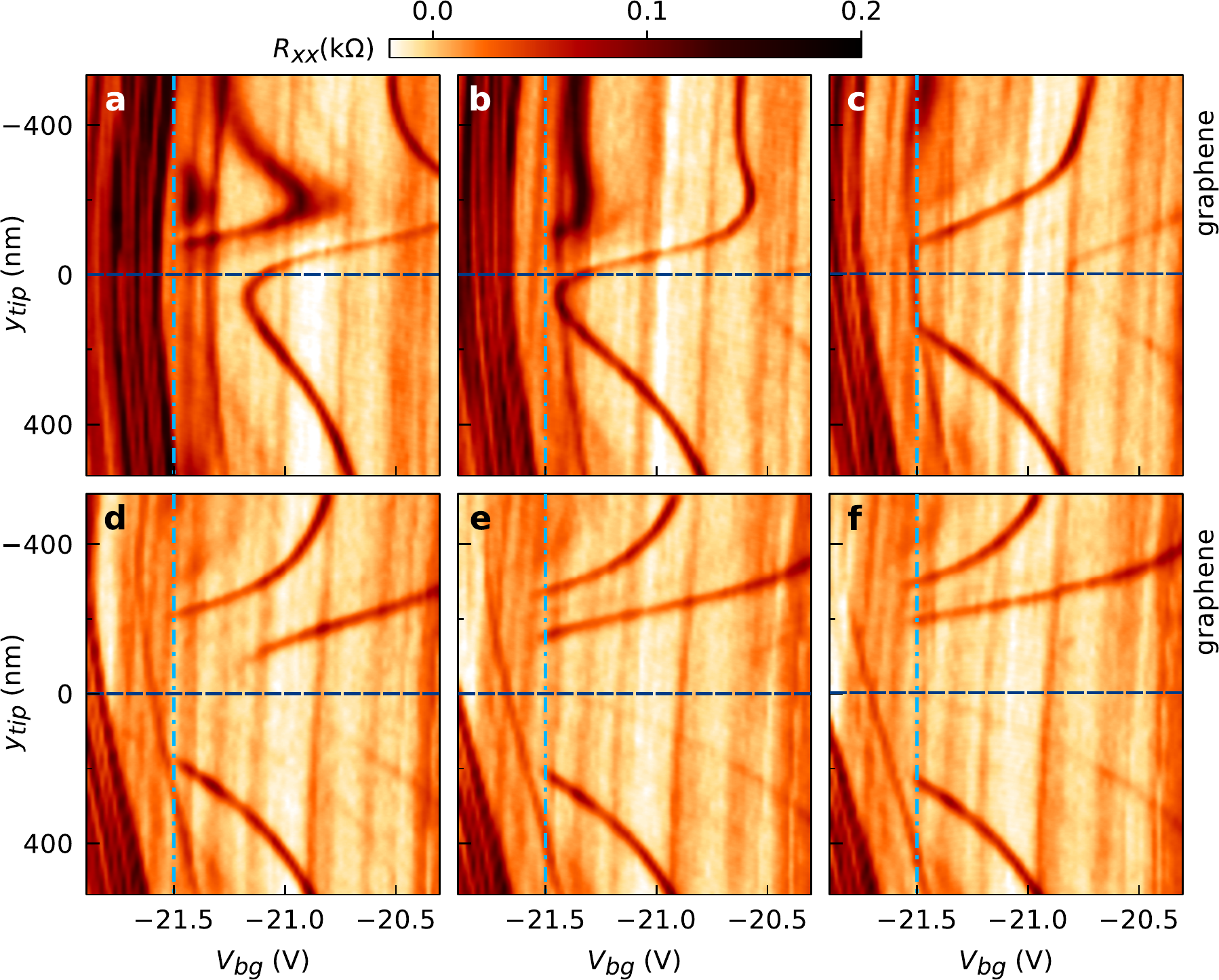}
\caption{\textbf{Effect of the screening on the tip-induced potential} \textbf{a-f}, $R_{xx}$ maps when varying $y_{tip}$ and $V_{bg}$ obtained in the same conditions as Fig. 3 of the manuscript with $V_{tip} = -3$ V (\textbf{a}), $-2$ V (\textbf{b}), $-1$ V (\textbf{c}), $0$ V (\textbf{d}), $+1$ V \textbf{e} and $+2$ V (\textbf{f}). Coulomb resonance signatures corresponding to the studied antidot all disappear when $V_{bg} < -21.5$ V (light blue dash-dotted lines), as discussed in the manuscript. The dark blue dashed lines indicate the device edge position.}
\label{tip_shapes}
\end{figure*}

Figure \ref{tip_shapes} gives the same maps as Fig. 3b of the manuscript for different values of $V_{tip}$. For each map, we expect Coulomb resonances to follow a Lorentzian evolution along $y_{tip}$ with an amplitude related to $V_{tip}$, as made in the manuscript. A striking feature however appears in Figs. \ref{tip_shapes}a,b. When following a Coulomb resonance line, it first reaches a maximum then a minimum value and the curvatures changes between them. In other words, the tip-induced potential exhibits a positive biased feature when the tip is outside the graphene plane and a negative biased feature when the tip is above graphene, far from the expected Lorentzian evolution. Schnez et al. observed the same kind of signature and attributed it to two different origins. First, the coupling between the tip and the studied antidot ($C_{tip}$ in Eq. \ref{charging}) is not symmetric on both sides of the antidot because of the difference of screening between the back gate and the graphene plane. Second, the coupling between the back gate and the antidot ($C_{bg}$ in Eq. \ref{charging}) also changes with the tip position due to screening effects. The combination of both effects explain how the features of Figs. \ref{tip_shapes}a,b emerge. 

As discussed in the manuscript, the resonance lines vanish below a given $V_{bg}$. In Fig. \ref{tip_shapes}, the $V_{bg}$ limit is the same in all maps and is indicated with blue dashed lines.

Finally, some resonances are also visible for $V_{bg} < -21.5$ V in Figs. \ref{tip_shapes} and in Fig. 3 of the manuscript. These resonances evolve slightly with the tip position. We ascribe them to signatures of the topological breakdown occurring between the $\nu = -6$ and $\nu = -10$ plateaus, far from the region scanned by the tip in Fig. 1 of the manuscript. It explains the small influence of the tip on these signatures.

\section{Coulomb diamonds}

\begin{figure*}[!ht]
\centering
\includegraphics[width=.8\linewidth]{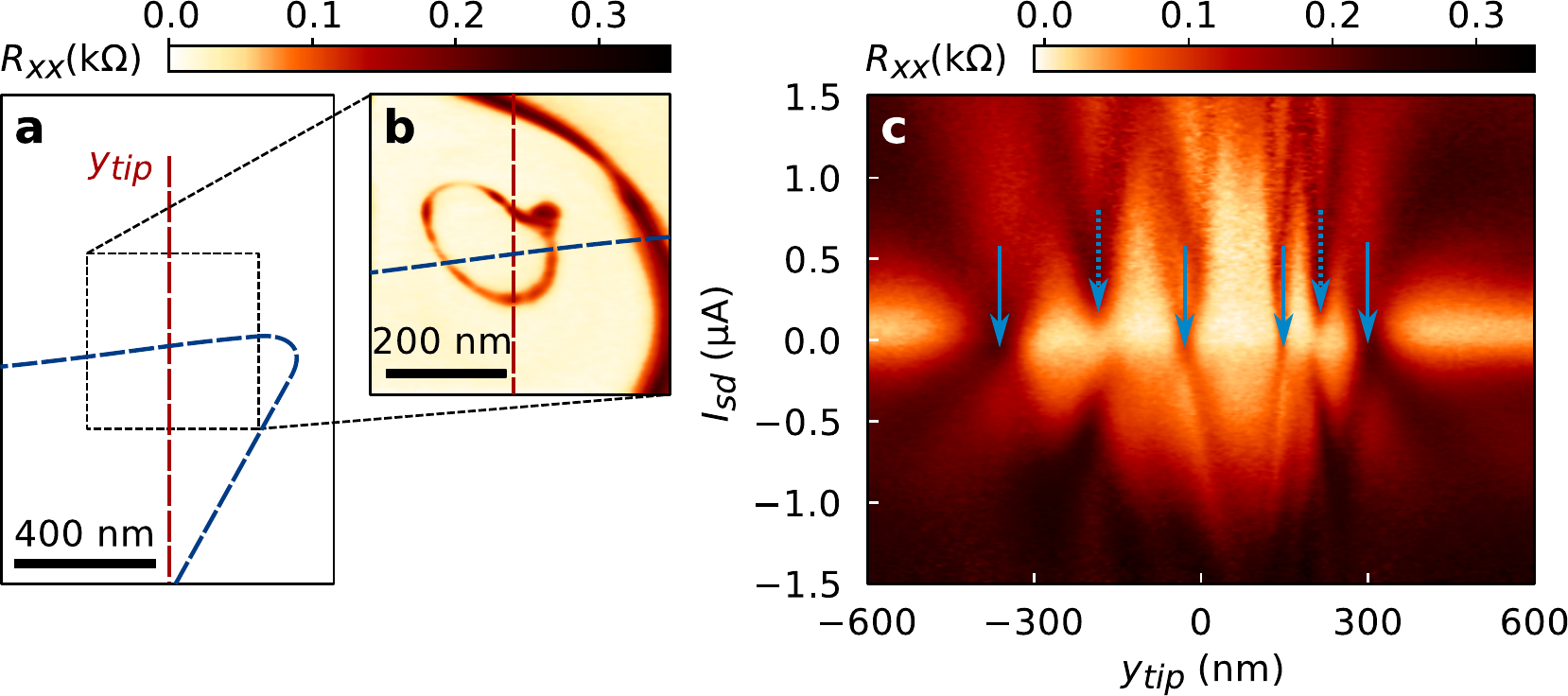}
\caption{\textbf{Coulomb diamonds related to a single antidot} \textbf{a}, SGM map obtained at 12 T with $V_{bg} = -17.75$ V and $V_{tip} = 0$ V in sample G1. The blue dashed line indicate sample edges. \textbf{b}, Zoom in the square area indicated with a dotted line in \textbf{a}. Coulomb rings are more contrasted due to slower scan rate. \textbf{c}, $R_{xx}$ map as a function of the tip position (along the red dashed line $y_{tip}$ in \textbf{a}) and the source-drain DC current $I_{sd}$. Blue arrow indicate the resonance signatures (Coulomb rings in \textbf{a} and \textbf{b}). These resonances widen by increasing the absolute value of $I_{sd}$ so that diamond-shaped regions of zero $R_{xx}$ appear in the graph. The dotted arrows indicate weaker resonances not visible in \textbf{a} and \textbf{b}.}
\label{Diamonds_line}
\end{figure*}

In this section, we show that the high $R_{xx}$ lines observed in our SGM data are indeed caused by transport through localized states (here associated with the presence of antidots). We drive this conclusion from Fig. \ref{Diamonds_line}c obtained by varying a DC bias voltage $V_{sd}$ (resulting in the measured DC current $I_{sd}$ through the sample) while moving the tip along the red dashed line in Figs. \ref{Diamonds_line}a,b. When applying a DC bias, the electrochemical potential levels of the input (source) and output (drain) leads are shifted with respect to each other and the energy window in which a discrete energy level associated with an antidot can contribute to charge carrier conduction is widened. It yields in the appearance of diamond shaped bright areas in the $R_{xx}$ map of Fig. \ref{Diamonds_line}c, indicating the couples of $V_{sd}$ and tip position for which transport through the antidot is forbidden (Coulomb blockade) so that counterpropagating QHECs are not coupled (no backscattering). These diamonds are separated by non-zero $R_{xx}$ regions corresponding to resonance signatures (highlighted with blue arrows). In that situation, charge carriers can flow through the antidot and counterpropagating QHECs are coupled.

\begin{figure*}[!ht]
\centering
\includegraphics[width=.8\linewidth]{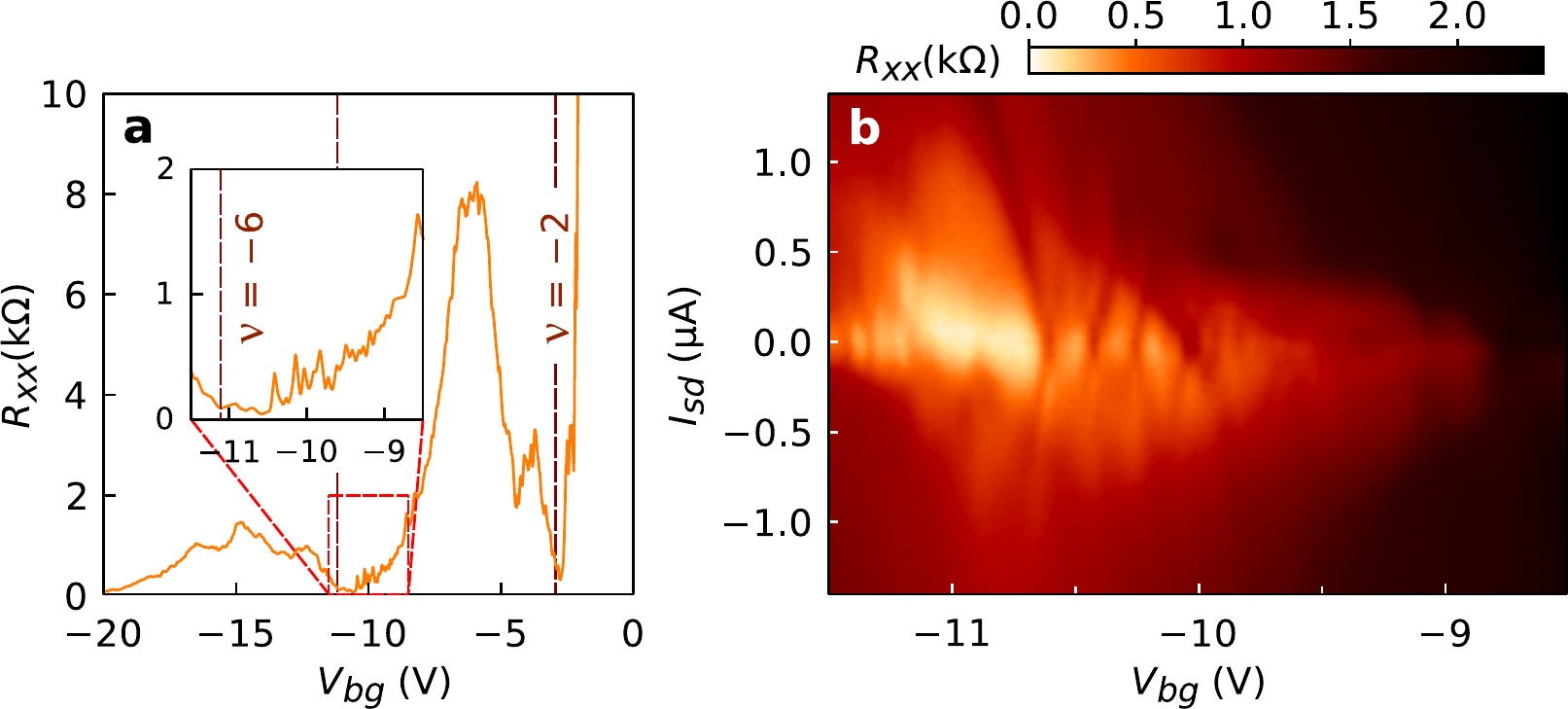}
\caption{\textbf{Coulomb diamonds obtained when varying back gate voltage} \textbf{a}, $R_{xx}$ measured as a function of $V_{bg}$ at 8 T in sample G1. The inset shows a zoom around $\nu = -6$ (red dashed line). $R_{xx}$ signal deviates from zero (loss of the QHECs topological protection) by exhibiting peaks. \textbf{b}, $R_{xx}$ map as a function of $V_{bg}$ and $I_{sd}$. Zero $R_{xx}$ Coulomb diamonds appear between each $R_{xx}$ peak of \textbf{a}.}
\label{Diamonds}
\end{figure*}

The same kind of low-$R_{xx}$-diamonds is obtained when varying both $I_{sd}$ and $V_{bg}$ as shown in Fig. \ref{Diamonds}b. Contrarily to Fig. \ref{Diamonds_line}c where the diamonds are associated to a single antidot\footnote{the resonance conditions are reached by changing the charging energy of this antidot only when moving the tip in its vicinity}, tuning $V_{bg}$ changes the charging energy of all the antidots in the system. The non-zero $R_{xx}$ regions between each diamond correspond to the $R_{xx}$ peaks observed at zero DC bias in the inset of Fig. \ref{Diamonds}a. It is therefore clear that these peaks are signatures of localized states resonances associated with antidots.

\section{Emergence of counterpropagating QHECs}\label{sec:pot}
In the manuscript, we show that the topological breakdown of QHECs is caused by the presence of both forward and backward QHECs along the same graphene edge and their coupling through antidots. In this section, we discuss the most widespread hypotheses about the appearance of these counterpropagating channels. We point out the different unsolved issues in these theories and the additional explorations that they required. 

In 2008, Silvestrov and Efetov \cite{Silvestrov2008} reported in a theoretical article that the screening of the back gate (modeled by an infinite conductor plane) by graphene charge carriers lead to the emergence of charge accumulation along the device edges. In the QH regime, it results in the appearance of counterpropagating QHECs along device borders. This theory has been widely invoked by several authors to explain experimental observations in the QH regime \cite{Vera-Marun2013, Barraud2014, Cui2016}.

In 2013, Vera-Marun et al. studied the evolution of the lever arm as a function of the distance from edges in graphene (and bilayer graphene) samples \cite{Vera-Marun2013}. To do so, they applied a perpendicular magnetic field that confined charge carriers along the device edges, in a region whose width is given by the cyclotron radius. Hence, they were able to probe the lever arm evolution between the bulk and the edges by changing the cyclotron radius of the charge carriers participating to conduction. Their results were close to theoretical simulations based exclusively on electrostatics, excepted when the cyclotron radius becomes smaller than 25 nm where the lever arm seamed to saturate. The lever arm increased by $\sim 100$\% between zero and high magnetic field.

In 2014, Barraud et al. conducted a similar study but found an increase of the lever arm of only 20\% between zero and maximal magnetic field \cite{Barraud2014}. They concluded that the charge density profile is not given solely by electrostatics. They claim that features depending on the studied sample, such as localized states at the edges, can reduce the theoretical charge accumulation due to electrostatics.

In 2016, Cui et al. observed that Hall conductance plateaus (corresponding to zero $R_{xx}$) don't necessarily coincide with an insulating bulk in graphene samples \cite{Cui2016} as expected in conventional QH effect observed in semiconductor-based two dimensional electron gas \cite{Lai2011}. To explain this surprising discovery, they use the charge accumulation model and the associated presence of counterpropagating channels in the QH regime. They state that QHECs region (where live forward and backward channels coexist in close proximity) is decoupled from the bulk by an incompressible (insulating) strip. Backscattering can therefore occur between counterpropagating channels along the same edge while the bulk is insulating or backscattering can be prohibited while the bulk is conducting. However, the shift they observed between Hall plateau and insulating bulk positions as a function of $V_{bg}$ is not always the same for holes and electrons. Their conclusion was the same as Ref. \cite{Barraud2014} : impurities at the edges locally modulate the electrostatic potential at the edges.

Finally, in 2019, Marguerite et al. \cite{Marguerite2019} used a SQUID-on-tip nanothermometer (that can also be used as a SGM) to study the breakdown of QHE topological protection in graphene. As in the present study, they observed signatures of large longitudinal resistance at the edges and invoked the presence of counterpropagating QHECs to explain their results. They however made a striking observation : at the charge neutrality point ($V_{bg} = 0$), signatures of counterpropagating QHECs persist. Inhomogeneous gating is not able to explain this observation since the increase of charge carriers density at the edges should be proportional to the applied $V_{bg}$ (here close to zero). They then conclude that a holes accumulation at graphene edges preexist in their samples, due to the presence of negatively charged impurities at the edges.

From these different studies, it appears that the exact shape as well as the physical origin of the inhomogeneous edge potential remain largely unknown. For our devices, the lever arm modulation is small when varying the magnetic field (the stripes of local resistance maxima only slightly deviate from the theoretical dashed straight lines in Figs. \ref{Fan_diagrams}a,c). This could be explained either by the presence of edge impurities affecting the potential modulation (Barraud et al. and Cui et al.) so that it saturates quickly near the physical device's borders or by the presence of a preexisting hole accumulation (Marguerite et al.). These two hypothetic mechanisms can furthermore coexist.

In conclusion, further studies are needed to characterize accurately the influence that edge impurities have on the potential modulation. In this work, we therefore use an arbitrary potential for simulation that exhibits qualitative features compatible with experimental evidences. That is, a bending of edge potential that quickly saturates when approaching device's borders (further details in section \ref{sec:simus}).

\section{Determination of the maximal tip-induced density change}\label{sec:tip}

In Fig. 4b of the manuscript, we converted the tip bias $V_{tip}$ into an induced change of carrier density. In particular, since we deal with hole carriers and we apply a positive tip bias, we induce a maximum decrease of the local hole density denoted $|\Delta n_{tip}|$. To convert $V_{tip}$ into $|\Delta n_{tip}|$, we can rely on the maps of Fig. \ref{tip_shapes}. The shift of the Coulomb resonances gives a direct correspondence between the effective tip-induced potential $\varphi_{tip}$ and $V_{bg}$ as a function of $y_{tip}$. Because the result of Fig. 4b were obtained with the tip above the antidot, $|\Delta n_{tip}|$ corresponds to the maximal change of density, obtained at the maximum of $\varphi_{tip}$. For each value of $V_{tip}$ of Fig. \ref{tip_shapes}, we then have
\begin{equation}\label{shift}
    \varphi_{tip}^{max}(V_{tip}) = \Delta V_{bg}(V_{tip})
\end{equation}
where $\Delta V_{bg}(V_{tip})$ is the difference between the maximal Coulomb resonance shift in $V_{bg}$ and the Coulomb resonance position in $V_{bg}$ without tip influence ($|y_{tip}| = \infty$). In Eq. \ref{shift}, the term $\varphi_{tip}^{max}(V_{tip})$ is supposed to be proportional to $V_{tip}$ \cite{Brun2019}. However, we must notice that $\Delta V_{bg}(V_{tip} = 0) \neq 0$ in Fig. \ref{tip_shapes}c (and Fig. 3b of the manuscript). This is due to the workfunction $V_{WF}$ between the tip and the graphene plane that adds an offset to the applied tip bias $V_{tip}$ so that 
\begin{equation}\label{workfunction}
    \Delta V_{bg}(V_{WF}) = 0
\end{equation}
The maximal tip-induced potential is then given by $\varphi_{tip}^{max}(V_{tip}) = \alpha_{tip}(V_{tip} - V_{WF})$ with 
\begin{equation}\label{lever_tip}
    \alpha_{tip} = \dfrac{\Delta V_{bg}(V_{tip})}{V_{tip} - V_{WF}}
\end{equation}
By using the result of Eq. \ref{lever_arm}, $|\Delta n_{tip}|$ can be obtained from Eq. \ref{shift} as
\begin{equation}\label{ntip}
    |\Delta n_{tip}| = \alpha|\Delta V_{bg}(V_{tip})|
\end{equation}
Based on Eq. \ref{ntip}, it is easy to link $|\Delta n_{tip}|$ to $V_{tip}$. 
From Eqs. \ref{ntip} and \ref{lever_tip}, we have
\begin{equation}\label{ntipVWF}
    |\Delta n_{tip}| = \left|\dfrac{\alpha}{\alpha_{tip}}\right|(V_{tip} - V_{WF})
\end{equation}
We must now assign values to the parameters $\alpha_{tip}$ and $V_{WF}$.

In section \ref{screening}, we discussed the central role of screening in the tip-induced potential experienced by an antidot located at the graphene device edge. We observed that both $C_{tip}$ and $C_{bg}$ changed as a function of $y_{tip}$, leading to the peculiar tip-potential evolution extracted in Fig. \ref{tip_shapes}a, at $V_{tip} = -3$ V. From this map, it appears that no value of $V_{tip}$ can exactly fulfill the condition of Eq. \ref{workfunction} since the effective tip-induced potential is different above graphene and outside the sample. We therefore had to estimate the work function value from the data of Fig. \ref{tip_shapes}. By took as criterion that the mean $\Delta V_{bg}$ along $y_{tip}$ should be close to 0. It led us to take $V_{WF} \sim -3$ V.

Finally, we estimated $\alpha_{tip}$ from Fig. 3b of the manuscript. In this figure, the Lorentzian fit gave us $\Delta V_{bg} = -1.55$ V. Considering the work function value determined just above, we obtained from Eq. \ref{lever_tip} that $\alpha_{tip} \sim 2$.

From these numerical values (such as the value of $\alpha$ given in section \ref{sec:lever-arm}), Eq. \ref{ntipVWF} gives us a direct link between the tip-induced decrease of holes density and the tip bias expressed as
\begin{equation}
    |\Delta n_{tip}| \sim (V_{tip} + 3) \times 4.7\sim 10^{14} ~ \text{m}^{-2}\text{V}^{-1}.
\end{equation}

\section{Measurements for electron-type charge carriers}

In the manuscript, such as in section \ref{screening}, all the presented results were obtained with hole type charge carriers. In this section, we explain why we were not able to lead a proper study with electron type charge carriers at high magnetic field in both studied samples.

In sample G1, it was not possible to properly define the longitudinal resistance $R_{xx}$ at high magnetic field for electron-type charge carriers, as shown in Fig. \ref{Fan_diagrams}a. Indeed, the current dropped to zero between $\nu = 0$ and $\nu = +4$, leading to diverging $R_{xx}$. It must be due to the loss of electrical contact between the metallic leads and the graphene plane. For $\nu > +4$, we recovered all the current we injected in the sample but the voltage $V_{xx}$ exhibited large chaotic fluctuations that we associate to poor equilibration of the metallic contacts with the QH QHECs. The SGM maps obtained with these parameters were therefore unexploitable. Nevertheless, we managed to extract data in the electron side at lower magnetic field that are discussed in \cite{Moreau2021}. In this paper, we discuss the role of contacts in the electron-hole asymmetry.

In sample G2, a leak between the electron gas and the back gate for $V_{bg} > +11$ V prevented us to make measurements with electron-type charge carriers.

\section{The different regimes of backscattering through an antidot}\label{backscattering}
In this section, we discuss in more details the different regimes of coupling between up- and downstream QHECs through an antidot. Indeed, the longitudinal resistance $R_{xx}$ reflects the backscattering of charge carriers tuned by the transmission $T$ between up- and downsteam QHECs. This transmission can be expressed as
\begin{equation}\label{trans}
    T = T_{da} T_{a} T_{au}
\end{equation}
where $T_{da}$ is the transmission between a downsteam QHEC and the antidot, $T_{a}$ is the transmission through the antidot and $T_{au}$ is the transmission between the antidot and the upstream QHEC. 

\begin{figure*}[!ht]
\centering
\includegraphics[width=\linewidth]{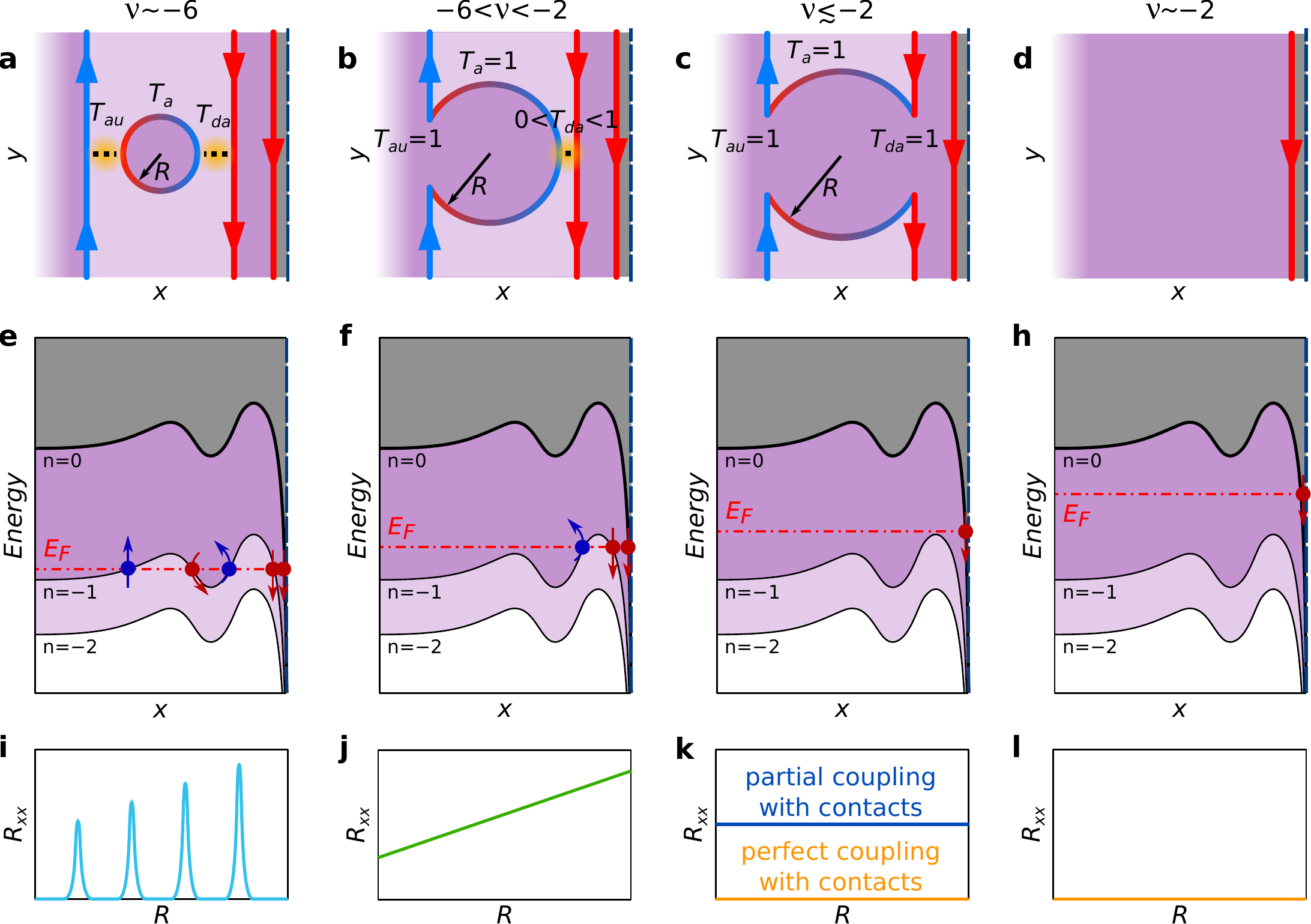}
\caption{\textbf{Different scenario of transmission between the up- and downstream QHECs through an antidot.} The downstream QHECs are represented in red and flow along the device edge (located downwards in each schematic). The upstream mode is in blue. The coupling between QHECs are represented with dotted lines and orange halos. \textbf{a}, The antidot is coupled to the upstream QHEC when the transmission $T_{au} > 0$ and to the nearest downstream QHEC when $T_{da} > 0$. The two downstream QHECs can be equilibrated when the transmission $T_{dd} > 0$. \textbf{b}, When the antidot is enlarged (by changing the potential in the vicinity of the antidot with the biased tip for instance), the QHEC loop of the antidot can be merged with the up- and downstream QHECs so that $T_{da} = T_{au} = 1$. If the two  downstream QHEC are not equilibrated at all ($T_{dd} = 0$), then $R_{xx}$ should indeed be zero. \textbf{c}, By considering that the potential is not symmetric on both sides of the antidot due to band bending (\textbf{f}), $T_{da} < T_{au}$. In this situation, while $T_{au} = 1$, the downstream QHEC is not fully equilibrated with the upstream QHEC and $R_{xx}$ is not zero. \textbf{d-f}, Schematics of the three first Landau levels in the vicinity of the antidot. Due to band bending, the potential is not symmetric on both sides of the antidot.}
\label{figure_Trans}
\end{figure*}

We now discuss three regimes linked to the radius $R$ of the antidot (see Fig. \ref{figure_Trans}). This radius is varied by changing either the Fermi energy $E_F$ compared to the Landau levels (through a change of the back-gate voltage $V_{bg}$), either through the local change of potential induced by the carge tip, as discussed in Fig. 4 of the manuscript.

In the first regime (Figs. \ref{figure_Trans}a,e), the antidot is tunnel-coupled to the down- and upstream QHECs. The corresponding transmission coefficients are respectively $0<T_{da}<1$ and $0<T_{au}<1$. In this case, the total transmission $T$ of Eq. \ref{trans} is given by the transmission $T_{a}$ in the antidot. As detailed in the manuscript, $T_a = 1$ when one of the discrete energy levels of the antidot is aligned with the Fermi level (Fig. 2c of the manuscript) and $T_a = 0$ otherwise (Fig. 2b of the manuscript). It results in a succession of peaks in $R_{xx}$ as the antidot's radius $R$ increases (Fig. \ref{figure_Trans}j). The maximal values of these peaks increases with $R$ since the transmissions $T_{da}$ and $T_{au}$ increases when the QHEC loop associated to the antidot draws nearer form the down- and upstream QHECs. 

In the second regime (Figs. \ref{figure_Trans}b,f), the antidot is so large that it merges with the upstream. Therefore $T_{au} = 1$. In this situation, the antidot is no more a closed system and does not carry discrete energies so that $T_a = 1$. The transmission $T$ only varies with $T_{da}$ that increases with $R$ (Fig. \ref{figure_Trans}j).

I the third regime (Figs. \ref{figure_Trans}c,g), the antidot still enlarges and merges with both the up- and downstream QHECs. Therefore, $T_{da} = T_a = T_{au} = 1$. In that case, the backscattering between down- and upstream QHECs is direct. In the case of a perfect coupling between the upstream QHEC and the injection contact, $R_{xx}$ is expected to vanish (Fig. \ref{figure_Trans}k, orange line). In this situation, the filling factor is given by the bulk and is therefore around $\nu = -2$. However, if the upstream QHEC is only partially coupled with the injection contact \cite{Moreau2021}, $R_{xx}$ remains finite (Fig. \ref{figure_Trans}k, blue line). This mechanism is a good candidate to explain the wide transition between two zero $R_{xx}$ regions in the $R_{xx}-V_{bg}$ curve in Fig. 1b of the manuscript and Fig. \ref{SGM_2}b. Finally, in the fourth regime (Figs. \ref{figure_Trans}d,h), the upstream QHEC disappears, such as the second downstream QHEC. $R_{xx}$ then vanishes (Fig. \ref{figure_Trans}l) and the filling factor is around $\nu = -2$.

Finally, in the fourth regime (Figs. \ref{figure_Trans}d,h), the upstream QHEC disappears, such as the second downstream QHEC. $R_{xx}$ then vanishes (Fig. \ref{figure_Trans}l) and the filling factor is around $\nu = -2$.

\section{Details on simulations}\label{sec:simus}

In this section, we detail the simulation methods. We first explain the choice of the tight-binding formalism to perform our simulation as well as the limitations inherent to this method. We then develop the tight-binding system used to reproduce the experimental results. We finish by presenting two different models that capture the main experimental features. 

\subsection{Tight-binding modelization and the absence of Coulomb energy}
As precised in the main text, simulations were performed in the tight-binding framework using the KWANT package \cite{Groth2014}. Whereas this package does not provide a recursive Poisson-Schr\"odinger solver, purely quantum mechanical-based calculations can capture the main physics we want to explore : coupling between QHECs via a localized state (an antidot). The major difference between simulations and the experiment therefore lays in the absence of Coulomb blockade in the antidot. Discrete energy levels originate only from the size confinement of the antidot (quantum terms $E_n$ in Eq. 1 of the main text). We therefore expect a qualitative correspondence between experiment and simulations, with the appearance of high $R_{xx}$ peaks when one of the discrete energy level aligns with the Fermi energy but we also expect a significantly quantitative difference in the energy spacing between two peaks, since Coulomb energy is neglected in simulations. This is indeed observed in Fig. 4 of the manuscript. 

\subsection{Construction of the tight-binding model}
In this section, we detail the tight-binding model we developed in order to simulate our experiment. Because the computation time increases fast with the number of atoms in the system \cite{Groth2014}, we can not simulate the whole sample. We therefore had to focus on a small region. 

\subsubsection{The tight-binding Hamiltonian}
The tight-binding Hamiltonian can be expressed in the second quantization formalism as \cite{Liu2015a}
\begin{equation}\label{hamiltonian}
    \hat{\mathcal{H}} = \sum_i U(\vec{r}_i)\hat{c}^{\dagger}_i\hat{c}_i - \sum_{<i, j>}t~ \hat{c}^{\dagger}_i\hat{c}_j 
\end{equation}
where $t = 2.7$ eV is the hopping parameter, $<i, j>$ is the restriction over the nearest neighboring atoms indices, $\hat{c}^{\dagger}_i$ and $\hat{c}_i$ are respectively the creation and annihilation operators on site $i$ and $U(\vec{r}_i)$ is the potential on site $i$. For an hexagonal lattice as graphene, the energy dispersion at low energies is linear as a function of the wavevector $\vec{k}$ such that
\begin{equation}
    E = \pm\dfrac{3at}{2}|\vec{k}|
\end{equation}
where $a = 1.42~\angstrom$ is the inter-atomic distance of graphene. In this expression, the constant ratio between $E$ and $|\vec{k}|$ is directly proportional to the Fermi velocity $v_F = 3at/2\hbar$. It is therefore possible to apply a scaling on $a$ to reduce the number of atom in the studied system while keeping both the Fermi energy $E_F$ and the Fermi wavelength $k_F$ of charge carriers unchanged. It implies to define the following scaled quantities \cite{Liu2015a}
\begin{equation}\label{scaling}
    a_{\text{scaled}} = a\times s \hspace{1cm} \text{and} \hspace{1cm} t_{\text{scaled}} = t/ s
\end{equation}
where $s$ is the scaling factor. 

The magnetic field $B$ is introduced in the tight-binding Hamiltonian of Eq. \ref{hamiltonian} by modifying the hopping parameter $t$ as
\begin{equation}
    t \longrightarrow t\exp \left( -j \frac{e}{h}B(x_i-x_j)\dfrac{y_i+y_j}{2} \right)
\end{equation}
where $(x_i, y_i)$ are the coordinates of site $i$. In graphene, the scaling transformations of Eq. \ref{scaling} also keeps LL energies unchanged while varying $s$ at constant magnetic field $B$. Indeed, these energies are given by \cite{Goerbig2011}
\begin{equation}
    E_{n} = v_F\sqrt{2\hbar e}\sqrt{nB}
\end{equation}
where $n = 0,~ \pm 1, ~ \pm 2,...$ is the LL label ($+$ for electrons and $-$ for holes). 

\begin{figure*}[!ht]
\centering
\includegraphics[width=.5\linewidth]{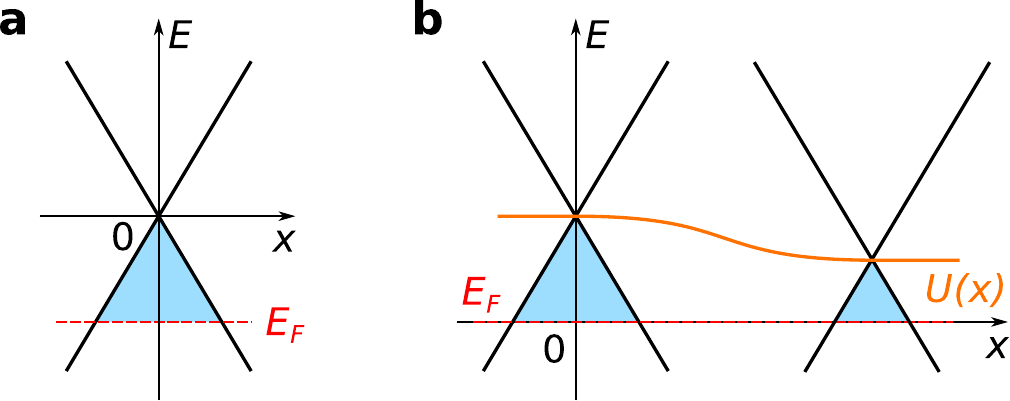}
\caption{\textbf{The density and the type of charge carriers can be defined in two equivalent ways.} \textbf{a}, Hole-type charge carriers are obtained by defining a negative $E_F$ with the onsite parameter $U = 0$ eV. \textbf{b}, Hole-type charge carriers are obtained by defining a positive onsite parameter with a zero Fermi energy $E_F = 0$ eV. Compared to the situation depicted in \textbf{a}, it has the advantage to easily change the charge carrier density in the space.}
\label{cones}
\end{figure*}

Since $k_F$ and $E_n$ does not change with the scaling factor $s$, the transport simulation outputs will not be affected by the choice of $s$ at a given $E_F$. Concerning the Fermi energy, we chose to change the onsite parameter $U(\vec{r}_i)$ in Eq. \ref{hamiltonian} while keeping $E_F = 0$ to adjust the charge carrier density. For a uniform potential, this is equivalent to adjusting the charge carrier density by changing $E_F$ with $U = 0$ eV. This is illustrated by Fig. \ref{cones}. To define hole-type charge carriers, the potential should be positive ($U>0$).

The tip-induced potential is modeled with a Lorentzian function added to the onsite term of Eq. \ref{hamiltonian}. The function is given by 
\begin{equation}
    V_{tip}(\vec{r}_i) = \dfrac{V_{max}}{1 + \left( \frac{\vec{r_i} - \vec{r}_{tip}}{R_{tip}} \right)^2}
\end{equation}
where $\vec{r}_i$ is the position of site $i$, $V_{max}$ is the tip potential maximum, $\vec{r}_{tip}$ is the tip position in the $x-y$ plane and $R_{tip}$ is half width at half maximum. From the experimental value (see Fig. 3b of the manuscript), we took $R_{tip} = 200$ nm (we didn't include the potential deformation due to screening). Here, $V_{max}$ is directly related to the tip-induced change of charge carriers density $\Delta n_{tip}$ discussed in section \ref{sec:tip} for the experiment. To compare the simulations and the experiment, we need to find a correspondence between the charge density and the energy.

The relation between the carrier density $n(\vec{r}_i)$ and the onsite parameter $U(\vec{r}_i)$ is simply given by \cite{Liu2015a}
\begin{equation}
    n(\vec{r}_i) = \dfrac{1}{\pi}\left(\dfrac{U(\vec{r}_i)}{\hbar v_F}\right)^2 
\end{equation}
By expressing the Fermi velocity in terms of $a$ and $t$ ($v_F = 3at/2\hbar$), $\Delta n_{tip}$ can easily be found by
\begin{equation}
    \Delta n_{tip} = \dfrac{4}{9\pi}\left(\dfrac{U_{bulk} + V_{max}}{at}\right)^2 - \dfrac{4}{9\pi}\left(\dfrac{U_{bulk}}{at}\right)^2
\end{equation}
where $U_{bulk}$ is the uniform onsite potential in the graphene bulk, as will be detailed in Fig. \ref{simu_square}d of the next section.

\subsubsection{Geometrical definition of the simulated system}

\begin{figure*}[!ht]
\centering
\includegraphics[width=\linewidth]{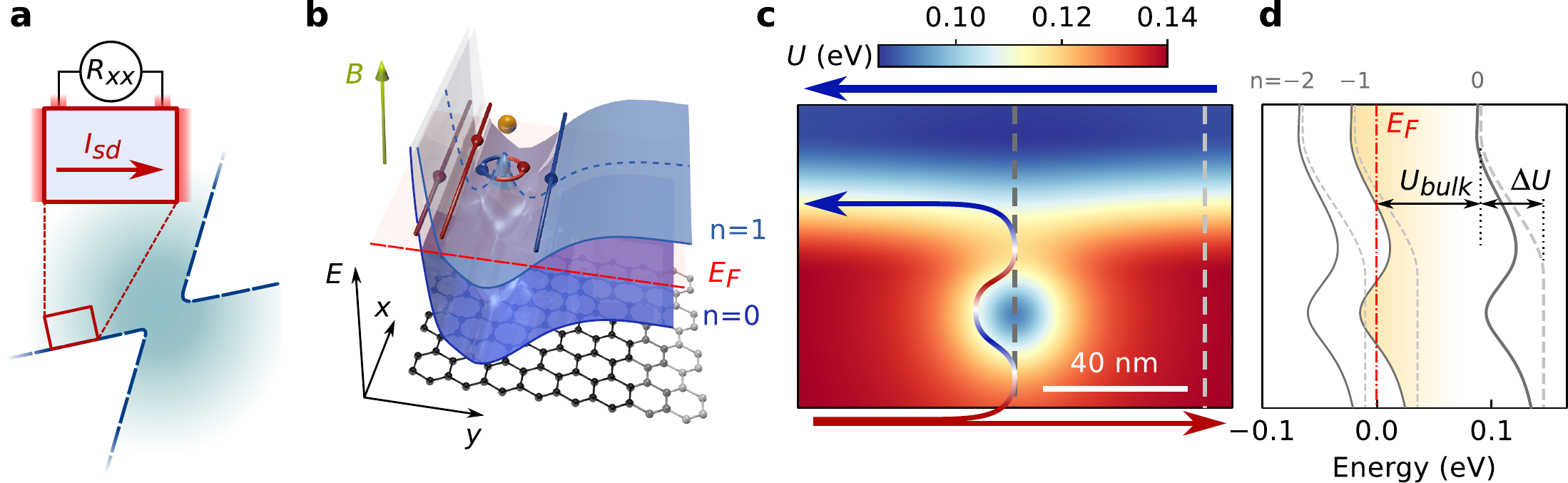}
\caption{\textbf{Definition of the simulated system} \textbf{a}, The simulated region is defined as the red rectangle. It is chosen to capture the physics along one of the edges of the whole sample used in the experiment. \textbf{b}, Artist's view of the physics we want to examine in the region defined in \textbf{a}. More details in Fig. 2a of the manuscript. \textbf{c}, Potential landscape (term $U(\vec{r}_i)$ in Eq. \ref{hamiltonian}), similar to Fig. 4e of the manuscript. The arrows illustrate the charge carriers flow in QHECs (forward in red and backward in blue). QHECs appear at the natural graphene edges (bottom red and top blue arrows). By considering only left to right transport, the backward top channel does not contribute. \textbf{d}, Three lowest Landau levels (LLs) along the dotted lines of \textbf{c}. The dark gray curves represent the LLs passing by the antidot and the light gray dashed curves represent the LLs that are not affect by the antidot potential. $U_{bulk}$ gives the potential value in the bulk and $\Delta U = \zeta U_{bulk}$ models the increase of charge density at the edges. Here, we chose arbitrarily $\zeta = 0.6$. The profiles in the potential $U$ of Fig. \textbf{c} are depicted with bold lines (coinciding with the $n = 0$ LL).}
\label{simu_square}
\end{figure*}

As discussed in detail in the manuscript, only the counterpropagating QHECs, located along the device edges, contribute to charge carriers transport, and the bulk is insulating. Furthermore, charge carriers flow without dissipation in forward propagating QHECs except when an antidot induces a coupling between forward and backward propagating channels (the coupling is determined by the transmission $T$ of Eq. (\ref{trans})). To capture the physics of the problem with simulations, we can therefore focus only on a small region along one of the device edges in which an antidot is located between the counterpropagating channels, as depicted in Fig. \ref{simu_square}a. A schematic view of the energy landscape with a high magnetic field applied perpendicularly to the studied region is presented in Fig. \ref{simu_square}b. Forward (red) and backward (blue) propagating QHECs appear at the intersection between the Fermi energy and the first Landau level (LL). An antidot (QH loop) is created due to a bump in the potential. 

In order to reach the situation depicted in Fig. \ref{simu_square}b, we designed the potential represented in Fig. \ref{simu_square}c and Fig. 4e in the manuscript. This potential corresponds to the onsite term $U(\vec{r}_i)$ in Eq. \ref{hamiltonian}. The potential profile shown in Fig. \ref{simu_square}d (or Fig. 4f in the manuscript) highlights the presence of an antidot located between the counterpropagating QHECs. The QHECs emerge at the crossing between the LLs and the Fermi energy $E_F$. In particular, the forward QHEC (red arrow at the bottom of Fig. \ref{simu_square}c) appears along the device border due to edge confinement (hard wall potential). In the same way, a "parasitic" backward channel (blue arrow at the top of Fig. \ref{simu_square}c) exists where the bulk is supposed to be insulating. We avoid the contribution from this QHEC by only computing transport from the left towards the right lead in the inset of Fig. \ref{simu_square}a. In this simulation, we ignore the coupling between QHECs and the ohmic contacts since we only concentrate on the coupling between up- and downstream QHEC through the antidot, determined by the transmission $T$ of Eq. (\ref{trans}). 

Simulations of the whole sample, including the study of the coupling between QHECs and the ohmic contacts, are presented in \cite{Moreau2021}. Nevertheless, the large size of the simulated sample in this study does not allow to have a precise characterization of the transmission through the antidot, that is perfectly captured in the system of Fig. \ref{simu_square}. 

\subsubsection{The different studied systems}
We simulated different systems to check that the observed results were resilient to changes of magnetic field, antidot size and modeling of the QHECs. Here, two systems of different size are detailed, including the system presented in the manuscript.

\begin{figure*}[!ht]
\centering
\includegraphics[width=.9\linewidth]{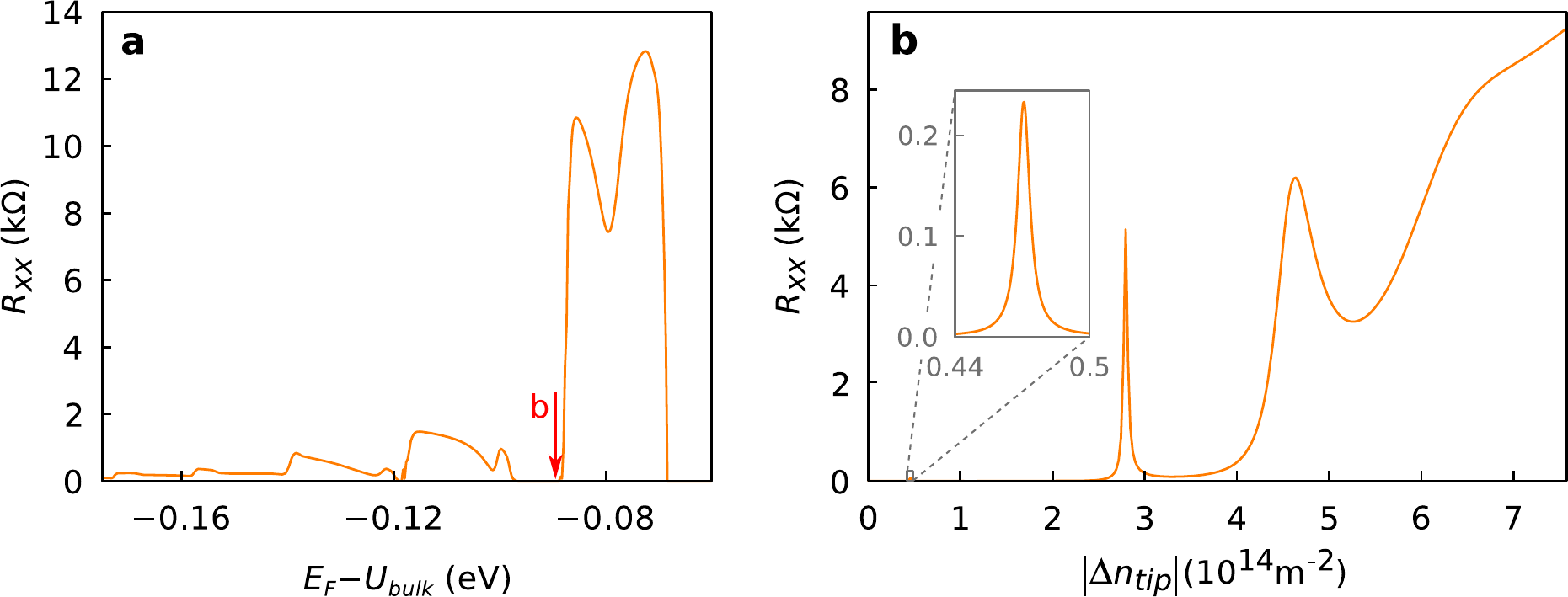}
\caption{\textbf{Transport measurements in the first studied system.} \textbf{a}, $R_{xx}$ as a function of $U_{bulk}$ (see Fig. \ref{simu_square}d). \textbf{b}, $R_{xx}$ as a function of the maximal tip-induced change of hole density $|\Delta n_{tip}|$ with $U_{bulk} = 0.0895$ eV (red arrow in \textbf{a}).}
\label{situation3}
\end{figure*}

The dimensions of the first studied system are $150\times 250$ nm$^2$ and the antidot is modeled with a Gaussian function centered $45$ nm away from the edge and having a FWHM of $72$ nm (see Fig. 4d of the manuscript). A scaling factor $s=4$ (Eq. \ref{scaling}) has been used to decrease the computation time. The $R_{xx}$ curve as a function of $U_{bulk}$ is shown in Fig. \ref{situation3}. The $R_{xx}$ curve as a function of $\Delta n_{tip}$ presented in Fig. \ref{situation3}b (same as Fig. 4c of the manuscript) is obtained for $U_{bulk} = 0.0895$ eV (red arrow in Fig. \ref{situation3}a). 

\begin{figure*}[!ht]
\centering
\includegraphics[width=.9\linewidth]{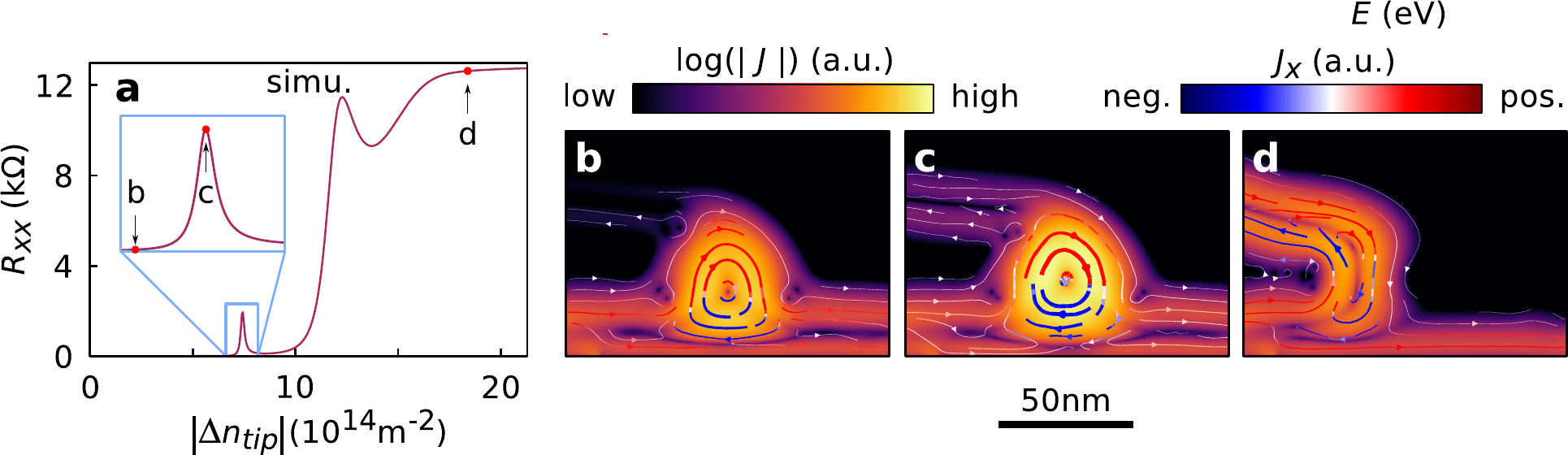}
\caption{\textbf{Transport measurements in the second studied system.} \textbf{a}, $R_{xx}$ as a function of the maximal tip-induced change of hole density $|\Delta n_{tip}|$ with $U_{bulk} = 0.089$ eV obtained with the potential presented in Fig. \ref{simu_square}c,d. \textbf{b-d}, Local current density maps obtained for the values $\Delta n_{tip}$ indicated with arrows in \textbf{a}.}
\label{situation2}
\end{figure*}

The second simulated system is presented in Figs. \ref{simu_square}c,d. The dimensions are $85\times 120$ nm$^2$ and the antidot is modeled by a Gaussian function of full width at half maximum (FWHM) of 28 nm and centered 26.5 nm from the edge. No scaling was applied to the inter-atomic distance ($a = 1.42~\angstrom$) and to the hopping parameter ($t = 2.7$ eV). $R_{xx}$ as a function of $\Delta n_{tip}$ obtained with a magnetic field of 12 T are presented in Fig. \ref{situation2}a. As for the first situation, $R_{xx}$ peaks appear by increasing the tip perturbation. However, the spacing between the peaks is wider than for the larger antidots (Fig. \ref{situation3}b). This is expected since the quantum discrete energy levels move apart from each others when decreasing the size of the localized state.

The local current density for different values of $\Delta n_{tip}$ are presented in Figs. \ref{situation2}b-e. As discussed in the manuscript, the current density in the antidot is higher on a $R_{xx}$ peak (Fig. \ref{situation2}c) than at zero $R_{xx}$ (Fig. \ref{situation2}b). It highlights that $R_{xx}$ peaks coincide with resonance conditions in the antidot. As detailed in Fig. 4i of the manuscript, the finite $R_{xx}$ plateau correspond to direct backscattering between the forward and the backward QHEC. 

\end{document}